\documentclass{pasj00}

\begin{document}
\SetRunningHead{Someya, Bamba, \& Ishida}{Progenitor Type Identification
of N103B}
\Received{2000/12/31}
\Accepted{2001/01/01}

\title{Progenitor Type Identification for 
Supernova Remnant N103B
in the Large Magellanic Cloud
by Suzaku and Chandra Observations}

\author{Kentaro \textsc{Someya}$^{1,2}$, 
Aya \textsc{Bamba}$^{3}$, and Manabu \textsc{Ishida}$^2$}
\affil{$^1$SoftBank Telecom Corp.,
1-9-1 Higashi-shimbashi, Minato-ku, Tokyo 105-7316, Japan}
\affil{$^2$Institute of Space and Astronautical Science, Japan Aerospace Exploration Agency, 3-1-1
  Yoshinodai, Chuo, Sagamihara, Kanagawa 252-5210, Japan}
\affil{$^3$Department of Physics and Mathematics, Aoyama Gakuin University,
5-10-1 Fuchinobe, Chuo, Sagamihara, Kanagawa 252-5258, Japan}
\email{bamba@phys.aoyama.ac.jp}


%


\maketitle

\begin{abstract}
This paper presents a detailed analysis of supernova remnant (SNR) N103B
located in the Large Magellanic Cloud (LMC), based on
Suzaku and Chandra observations.
The spectrum of the entire remnant was closely reproduced
using three interstellar medium (ISM) components with temperatures of
$\sim$0.32, $\sim$0.56, and $\sim$0.92~keV
and one ejecta component with a temperature of $\sim$3.96~keV,
based on spectral analysis of the Suzaku/X-ray Imaging Spectrometer (XIS) data.
The ejecta was overabundant in heavy elements, such as
Mg, Si, S, Ca, Fe, and Ni.
The unprecedentedly high quality of data obtained by Suzaku/XIS,
allowed us to correctly distinguish between the emissions from the ISM and the ejecta
 for the first time in a spectral analysis.
Combining spectral analysis based on Suzaku/XIS data with image analysis
based on Chandra/Advanced CCD Imaging Spectrometer (ACIS) data,
we verified that the ejecta distributions for elements from Si to Fe-K were
similar to one another,
although Fe-K emission was located slightly inward
compared with that of lighter elements such as Si, S, Ar, and Ca.
The onion-like structure of the ejecta was maintained
after the supernova explosion.
In addition, the ISM emission represented
by O and Fe-L was located
inside the ejecta emission.
We compared hydrogen-rich ejecta plasma
(called ``H-dominated plasma''),
which is indicative of Type II SNRs,
with plasma rich in heavy elements and poor in hydrogen
(called ``pure metal plasma''), which is
mainly observed in Type Ia SNRs.
In the case of N103B,
we could not determine whether the origin of the continuum emission
in the 4.0--6.0~keV band
was from ejecta (H-dominated plasma)
or high-temperature ISM (pure metal plasma)
only based on the spectral modeling of Suzaku/XIS data.
High-energy continuum images in the 5.2--6.0~keV band 
obtained by Chandra/ACIS were extremely similar to those of ejecta,
implying that
the origin of the high-energy continuum might indeed be the ejecta.
By combining spectral analysis with high-energy continuum images,
we found some indications for H-dominated plasma,
and as a result, that the progenitor of N103B might have been
a Type II supernova.
The progenitor mass was estimated to be 13~M$_\odot$
based on the abundance patterns of Mg, Fe, and Ni relative to Si.
\end{abstract}

\section{Introduction}

A major challenge in modern astronomy is to elucidate the chemical evolution in the universe.
Heavy elements are distributed
throughout the universe mainly by supernova explosions, which are classified
into two basic types: core collapse (CC) of massive stars
(Type Ib, Ic, and II)
and thermonuclear runaway of white dwarfs (Type Ia).
Supernova explosions play an important role in chemical evolution
being responsible for the distribution of heavy elements
inside a galaxy,
yielding the O group in the case of CC and the Fe group in the case of Type Ia.
Accordingly, studying supernova remnants (SNRs) is essential.
Typical SNRs are bright sources in the X-ray band
due to the presence of high-temperature plasma ($\sim10^{7}$~K) heated by the blast wave.
Through X-ray observations, we can gather information about the progenitor
from the ejecta heated by the reverse shock
as well as information about the surrounding environment from ISM heated by the forward shock.
Such information is expected to provide insights into galactic chemical evolution
and star formation history.

During SNR evolution, the
X-ray emission from a young SNR at the start of the Sedov phase is complex
since it is composed of emission from both ISM and ejecta.
To estimate physical parameters such as plasma ionization age,
number density, and plasma abundance,
we need to distinguish between emissions from ISM and ejecta through image and/or spectral analysis.
However, the number of young SNRs for which we can separate these components
is limited.
Currently, several in-orbit X-ray satellites
possess high-quality spectroscopic capabilities and large effective area,
(such as Suzaku and XMM-Newton)
or superior imaging capabilities (such as Chandra).
These satellites
allow us to study complex emissions from young SNRs.

The distance to the Large Magellanic Cloud (LMC) is  48~kpc \citep{2006ApJ...652.1133M},
and interstellar absorption for SNRs in the LMC
has a relatively small column density ($\sim10^{21}~\rm{H\,cm^2}$)
compared with most Galactic SNRs.
These properties are advantageous for estimating
the physical parameters of the plasma and determining the progenitor type.
The growing abundance of high-quality observation data provided by Suzaku,
XMM-Newton, and Chandra
warrants a re-examination of research on SNRs in the LMC.
In this paper, we present the results of a study of the bright young SNR N103B in the LMC.
N103B is one of the bright radio and X-ray SNRs in the LMC, and it was identified as an SNR based on its non-thermal spectrum
in radio observations \citep{1983ApJS...51..345M}.
\citet{2005Natur.438.1132R} estimated the age of N103B to be 860~yr
based on the light echo, making this remnant one of the younger SNRs in the LMC.
Despite being the subject of many X-ray observations
(e.g., ASCA \citep{1995ApJ...444L..81H},
XMM-Newton \citep{2002A&A...392..955V},
and Chandra \citep{2003ApJ...582..770L}),
the progenitor type and emission origin for this remnant
are not well understood.
\citet{1988AJ.....96.1874C} reported that there is the HII region DEM~L84
and the young rich cluster NGC1850 close to N103B,
which suggest a CC origin.
Observations with the Reflection Grating Spectrometer (RGS)
onboard XMM-Newton by \citet{2002A&A...392..955V}
also indicated that the origin might be CC, based on the O, Ne, and Mg emissions.
Nevertheless, \citet{2009ApJ...700..727B} suggested that
N103B might have had a relatively younger and more massive Type Ia progenitor
that underwent substantial mass loss before the explosion.
\citet{2003ApJ...582..770L} also suggested a Type Ia progenitor
based on the large mass corresponding to Fe.
In a recent observation, \citet{2011ApJ...732..114L} conducted a statistical image analysis and reported that
the progenitor was Type Ia based on the morphologies
of Si~XIII and 0.5--2.1~keV band emissions
.
\citet{2013ApJ...766...44Y} also suggested a Type Ia progenitor
based on the Cr/Mn ratio.
The origin of the emission is also unclear.
For instance, the O emission has been attributed to ambient material \citep{2003ApJ...582..770L}
and ejecta \citep{2002A&A...392..955V}.

In this paper, we present the X-ray spectrum and images of N103B obtained 
with Suzaku and Chandra, respectively.
Detailed observations and data reduction are described in \S~\ref{sec:obs}.
Spectral analysis results are summarized in \S~\ref{sec:spe}.
Owing to the high-quality spectra acquired by Suzaku,
we were able to distinguish between emissions from ISM and ejecta,
and successfully determined the abundances for the first time in a spectral analysis.
In \S\ref{sec:img}, we present image analysis performed using data from Chandra.
On the basis of these results,
we discuss the plasma parameter and the progenitor type in \S\ref{sec:dis}.
A summary is given in \S\ref{sec:sum}.

\section{Observation \& Data Reduction}\label{sec:obs}

\begin{table*}
  \caption{Observation log for N103B.}\label{tab:obslog}
  \begin{center}
    \begin{tabular}{lllll}
      \hline
& Suzaku & Chandra \\ \hline
     Observation ID \dotfill & 100013010 & 1045 \\
     Observation mode \dotfill & Full window, no burst, no SCI & FAINT \\ 
     Exposure (ks) \dotfill & 33 & 41 \\
     Count rate (counts~s$^{-1}$)\dotfill & FI: 2.1$^*$ BI: 3.7$^*$ & 0.6$^\star$ \\
      \hline
    \end{tabular}
  \end{center}
 {$^*$ Count rates of FI- and BI-CCDs of N103B
in the 0.3--10.0~keV and 0.55--10.0~keV bands
within a circle of radius $\sim$4.3' after background subtraction,
respectively.}\\
 {$^\star$ Total count rate within a circle of radius 3'
after background subtraction.}
\end{table*}


\subsection{Suzaku}

N103B was observed by Suzaku \citep{2007PASJ...59S...1M}
on August 30--31, 2005, as one of the Science Working Group targets.
The observation log is summarized in Table~\ref{tab:obslog}.
Suzaku is equipped with four X-ray CCD cameras
(X-ray Imaging Spectrometer (XIS) 0, 1, 2, and 3: \citet{2007PASJ...59S..23K})
in the focal planes of the four X-ray telescopes (XRTs: \citet{2007PASJ...59S...9S}),
whose half-power diameters are $\sim$1.8'--2.3'.
XIS 0, 2, and 3 are equipped with front-illuminated (FI) CCDs with high sensitivity
in the high energy band,
whereas XIS 1 is a back-illuminated (BI) CCD
with superior sensitivity and spectral resolution
at low energies (below $\sim$1~keV).
XIS instruments were operated
in normal full-frame clocking mode without
spaced-row charge injection \citep{nakajima2008,uchiyama2009}.

We analyzed the data taken in the $5\times5$ and $3\times3$ editing mode
with revision 2.0 of the {\sc HEAdas} 6.5.1 software package.
In screening the data, we removed the time intervals
corresponding to the South Atlantic Anomaly and night-earth and day-earth elevation angles
 less than 5$^\circ$ and 20$^\circ$, respectively.
We utilized only events with a grade of 0, 2, 3, 4, and 6
in the following spectral analysis.
The total exposure was $\sim$33~ks.

\subsection{Chandra}

The spatial structure of this remnant
cannot be examined reliably with the image performance of Suzaku/XRT,
whose half-power diameter is $\sim2'$, since N103B has a small radius of $\sim15$''.
We therefore performed image analysis using Chandra data
to facilitate the spectral analysis of the Suzaku/XIS data
for the entire remnant.
N103B (ObsID 1045) was observed
with the Advanced CCD Imaging Spectrometer (ACIS) onboard Chandra
on December 4, 1999 (see also \citet{2003ApJ...582..770L}).
The observation was conducted in {\sc faint} mode.
We began our analysis from Level 2 event files processed
with calibration data in CALDB version 3.4.1.
The observation data were not contaminated by background flares,
and therefore we used the full exposure of $\sim$41~ks.
In the following image analysis,
narrow-band images were extracted using the {\sc dmcopy} command
in the Chandra Interactive Analysis of Observations (CIAO) software package version 4.1.2, and we verified the nominal energy column
for the energy selection (see also \citet{2002A&A...392..955V}).

\section{Spectral Analysis Using Suzaku/XIS Data}\label{sec:spe}

\subsection{Extraction of the Source Photons}

\begin{figure}
  \begin{center}
   \FigureFile(80mm,80mm)
   {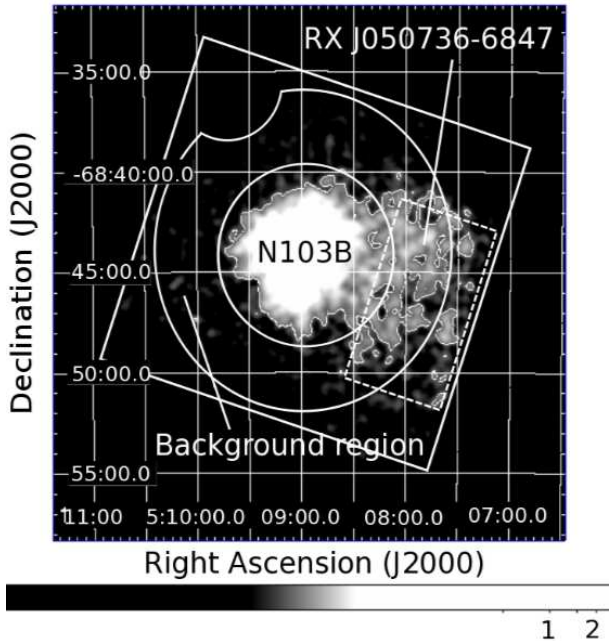}
  \end{center}
  \caption{Suzaku/XIS image of N103B in the 0.2--0.7~keV band.
The images from all four XIS modules were combined.
The white solid square shows the XIS field of view.
The $^{55}$Fe calibration sources in each XIS corner were masked.
The image was smoothed using a Gaussian function with a standard deviation of $\sim$21''.
The coordinates are represented in the equatorial coordinate system
for epoch 2000.
The contour was overlaid with a level of 0.02 counts px$^{-1}$.
The extended source to the west is
a known SNR, RX~J050736$-$6847 \citep{2000AJ....119.2242C}.
The white solid circle and annulus represent
the source and background regions of N103B, respectively,
and the white dotted region is the source RX J050736-6847.}
 \label{fig:xisA_N103B_200-700eV}
\end{figure}

Figure~\ref{fig:xisA_N103B_200-700eV} shows a Suzaku/XIS image
in the 0.2--0.7~keV band.
The centered source is SNR N103B,
and the extended source to the west is SNR RX~J050736$-$6847
\citep{2000AJ....119.2242C}, as observed by ROSAT.
According to \citet{2000AJ....119.2242C},
the position and radius of RX~J050736$-$6847 are
($\alpha$, $\delta$) = (05:07:36, -68$^\circ$47'52") and $\sim5'$,
respectively, corresponding to $\sim$70~pc in the LMC at a distance of 48~kpc
\citep{2006ApJ...652.1133M}.

To extract the source photons for N103B,
we selected a circular region with a radius of $\sim4.3'$ around N103B
as the source region,
while an annular region with an outer radius of $\sim7.8'$
around the source region was selected as the background region.
The $^{55}$Fe calibration sources in the corner of the each XIS were masked.
Figure~\ref{fig:xisA_N103B_200-700eV} shows that
the X-ray emission from RX~J050736$-$6847 contaminates
both the source and background regions for N103B.
We therefore selected a rectangular region ($\sim5.7'\times9.4'$)without the N103B source region
to estimate the contamination from RX~J050736$-$6847.
For the background region of this estimation,
we used the same annular region
excluding the RX~J050736$-$6847 source region.
As a result,
the background-subtracted count rates of FI- and BI-CCDs
from RX~J050736$-$6847.8 were roughly 0.02 and 0.06~counts\,s$^{-1}$
in the 0.5--2.0~keV band, respectively.
We therefore ignored the emission from RX~J050736$-$6847.8
in the following spectral analysis
since these count rates amount to only $\sim$1\%--2\%
of those from N103B (see Table~\ref{tab:obslog}).

In the spectral analysis,
we used {\sc xspec} version 11.3.2.aj.
The XIS response matrix files (RMFs) and auxiliary response files (ARFs) were
prepared with {\sc xisrmfgen} version 2007-05-14
and {\sc xissimarfgen} version 2008-04-05, respectively.
The prepared ARFs and RMFs were combined with
{\sc addrmf} and {\sc marfrmf} in the FTOOLS package.

\subsection{Full-band Spectra}\label{sec:H_dominated}

\subsubsection{Spectral Model}

Based on observations by XMM-Newton \citep{2002A&A...392..955V}
and Chandra \citep{2003ApJ...582..770L},
the emission was found to be closely reproduced
by the optically thin thermal emission model
under non-equilibrium ionization.
We therefore adopted a {\sc vnei} model
in the {\sc xspec} library (\cite{1983ApJS...51..115H},
\cite{1994ApJ...429..710B}, \cite{1995ApJ...438L.115L},
\cite{2001ApJ...548..820B}).
In the spectral analysis,
we obtained the plasma temperature ($kT_{\rm{e}}$),
the metal abundance $A_{\rm{i}}$
(relative to the solar abundance),
the ionization parameters $n_{\rm{e}}t_{\rm{ion}}$
(where $n_{\rm{e}}$ and $t_{\rm{ion}}$ are the plasma number density
and the elapsed time of the shock, respectively),
and a normalization $\frac{10^{-14}}{4\pi D^2}\int n_{\rm{e}}n_{\rm{H}}dV$
(where $D$, $n_{\rm{H}}$, and $dV$ are the distance to the source,
the hydrogen number density, and the volume element of plasma, respectively).
It should be noted that the {\sc vnei} code version 2.0 does not include K-shell emission lines
for ions below the He-like state.
We therefore used {\sc vnei} code version 1.1, which in turn does not include the line emission from Ar., and we specifically added a {\sc Gaussian} model to the {\sc xspec} library
for the Ar He-like K$\alpha$ transition ($\sim$3.14~keV). 

The average metal abundances in the LMC have been measured to be $\sim -$0.3~dex
\citep{1992ApJ...384..508R}.
We calculated the absolute abundances using the data presented by \citet{1992ApJ...384..508R}.
Since we used the abundance table from \citet{1989GeCoA..53..197A}
in {\sc xspec},
we compared the calculated abundances with those in \citet{1989GeCoA..53..197A}
and derived the relative abundance for {\sc xspec}.
The best value in the second column of Table~\ref{table:abs_abundances} shows
the relative abundances used in this study.
For Si, we used the value by \citet{1992ApJ...384..508R}
(right column of Table~\ref{table:abs_abundances}.
These values are consistent with those 
in other studies
\citep{1998ApJ...505..732H,2006ApJ...642..834K,2010A&A...517A..50G}.

\begin{table*}
  \caption{Metal abundances in the LMC and comparison with other observations.}
  \label{table:abs_abundances}
  \begin{center}
    \begin{tabular}{ccc}
      \hline
	Element & \citet{1992ApJ...384..508R}$^*$ & \citet{1998ApJ...505..732H}$^\star$ \\ \hline
	O   & 0.26 (0.22--0.30) & 0.19 (0.16--0.32)\\
	Ne & 0.33 (0.29--0.37) & 0.29 (0.24--0.35)\\
	Mg & 0.78 (0.58--1.05) & 0.32 (0.27--0.37)\\
	Si   & 1.82 (highly uncertain) & 0.31 (0.26--0.37)\\
	S    &  0.31 (0.25--0.38) & 0.36 (0.27--0.49)\\
	Ca &  0.34 (0.23--0.49) & -  \\
	Fe  & 0.36 (0.26--0.50) & 0.22 (0.17--0.28) \\
	Ni   & 0.62 (0.50--0.76) & - \\
      \hline
  \multicolumn{3}{l}{$^*$Si abundance is highly uncertain.}\\
\multicolumn{3}{l}{$^\star$Ca and Ni abundances have not been measured.}
    \end{tabular}
  \end{center}
\end{table*}

Emission from sources in the LMC is attenuated
through absorption by interstellar matter in both the Milky Way and the LMC.
Since the metal abundances of these two absorption components are
significantly different,
we estimated the absorptions as follows.
The abundances in absorbing matter in our Galaxy
were assumed to be solar abundances \citep{1989GeCoA..53..197A},
and the hydrogen column density was denoted as $N_{\rm{H}}^{\rm{Galactic}}$.
The value of $N_{\rm{H}}^{\rm{Galactic}}$ for N103B
has already been determined to be $\sim6.2\times10^{20}~\rm{H\,cm^{-2}}$  in the Galactic HI survey
\citep{1990ARA&A..28..215D}.
We fixed $N_{\rm{H}}^{\rm{Galactic}}$ at this value
and used the {\sc phabs} model in the {\sc xspec} library.
We also assumed that the absorbing matter in the LMC has the abundance pattern 
shown in Table~\ref{table:abs_abundances}.
This absorption component is denoted as $N_{\rm{H}}^{\rm{LMC}}$,
and we used the {\sc vphabs} model in the {\sc xspec} library,
and allowed the value to vary.

The calibration uncertainty of the normalization
and the hydrogen column density for XIS 0, 1, 2, and 3 were
estimated to be roughly $\sim$1\%--10\%
and (3.0--3.5)$\times10^{21}~\rm{H\,cm^{2}}$, respectively,
from the estimation of the systematic uncertainty for the Crab Nebula
\citep{2007PASJ...59S...9S}.
We therefore allowed these parameters to fluctuate
between FI and BI CCDs in the following spectral analysis.
In addition, in the early phase of the Suzaku mission,
the energy gain of XIS had an uncertainty of a few eV
\citep{2007PASJ...59S..23K}.
In previous studies on SNRs,
the uncertainty was estimated to be $\sim$10~eV at a maximum 
(see also \citet{2008PASJ...60S.141Y}, \citet{2008PASJ...60S.153B},
and \citet{2010PASJ...62.1301S}).
We used the range of 0.55--10.0~keV for FI and 0.3--10.0~keV for BI,
and excluded data in the 1.7--1.8~keV band
due to the uncertainty of XIS calibration around the Si-K edge
\citep{2007PASJ...59S..23K} in the spectral analysis.

The emission of SNRs is a mix of
what is produced in different plasma components
in the ejecta and the ISM.
In some cases, especially for young SNRs,
this is natural since
both ISM and ejecta can be non-uniform,
and the shock wave propagates with a  long time scale to heat them up.
Performing fitting with such a complex model
is associated with considerable uncertainty in determining the parameters.
Therefore, we began with a simple plasma model
and added more free parameters if necessary.

\begin{table*}
  \caption{Plasma models and their respective $\chi^2/d.o.f.$ values.}
  \label{table:chi}
  \begin{center}
    \begin{tabular}{clr}
      \hline
      	Model$^*$ &  \multicolumn{1}{c}{Plasma origin$^\dagger$} & \multicolumn{1}{c}{$\chi^2 / d.o.f.$ $(= \chi^2_{\nu})$} \\\hline
	1~{\sc vnei} component & (a) One-temperature plasma & 5102.99/ 471 (= 10.8) \\
	2~{\sc vnei} components & (b) Two-temperature plasma & 1424.11/458 (= 3.11) \\
	3~{\sc vnei} components & (c) 3ISM or 3Ejecta & 1267.236/465 (= 2.73) \\
	3~{\sc vnei} components & (d) 1ISM + 2Ejecta  & 1062.649/463 (= 2.30) \\
	3~{\sc vnei} components & (e) 2ISM + 1Ejecta  &   820.264/463 (= 1.77) \\
	4~{\sc vnei} components & (f) 2ISM + 2Ejecta   &   795.709/460 (= 1.73)\\
	4~{\sc vnei} components & (g) 3ISM + 1Ejecta  &   632.837/460 (= 1.38)\\
	5~{\sc vnei} components & (h) 3ISM + 2Ejecta  &    626.858/457 (= 1.37)\\
	5~{\sc vnei} components & (i) 4ISM + 1Ejecta   &    645.656/457 (= 1.38) \\	
      \hline
    \end{tabular}
  \end{center}
     {$^*$ Indicates the number of {\sc vnei} components.}\\
     {$^\dagger$ Model (a) assumes one-temperature ISM
or ejecta, whereas
model (b) assumes two-temperature ISM,  two-temperature ejecta,
or one-temperature ISM with one-temperature ejecta.
In models (c) -- (i), ``ISM'' and ``Ejecta'' indicate
interstellar matter and ejecta components, respectively.
The number $x$ in $x$ISM and $x$Ejecta represents
the number of components of the respective type.
We took the metal abundance to be the same for
all multi-temperature variable abundance non-equilibrium ionization ({\sc vnei}) components of the same plasma origin.
For instance, the 1ISM + 2Ejecta model represents
emission from one ISM component
and two ejecta components.
In the assumption of ISM and ejecta,
the abundances, which could not be determined
from spectral analysis, were fixed
at the LMC average \citep{1992ApJ...384..508R} and 1~solar, respectively.}
\end{table*}

\subsubsection{Models with One and Two Plasma Components}

We began with the spectral analysis
with models using 1 or 2 {\sc vnei} components
attenuated by photoelectric absorption.
The abundances of elements from C to Ni were set to vary in each plasma
to maximize the degree of freedom.
These models were unable to provide an acceptable fit.
The $\chi^2/d.o.f.$ values for these models are summarized
in Table~\ref{table:chi} under (a) One-temperature plasma
and (b) Two-temperature plasma.

\subsubsection{Model with Three Plasma Components}\label{sec:tts}

In the previous section,
we found that a reasonable fit of the emission requires
at least three plasma components.
We therefore adopted a third {\sc vnei} component.
The European Photon Imaging Camera (EPIC) Metal-Oxide Semiconductor (MOS)
and RGS onboard XMM-Newton observations
\citep{2002A&A...392..955V} have already demonstrated that
we need at least three thermal components.
To minimize the number of model parameters,
we assumed that
some plasma components have a common origin,
and show a common abundance pattern.
Thus, ISM should have
abundance patterns similar to that of LMC,
whereas the abundance would be greater for ejecta.
We divided the fitting into 3 cases, namely
(c) 3ISM or 3Ejecta, (d) 1ISM + 2Ejecta and (e) 2ISM + 1Ejecta,
where ``ISM'' and ``Ejecta'' denote interstellar medium
heated by the forward shock and ejecta heated by the reverse shock,
respectively.
From model (c),
we assumed the metal abundances to be the same
for all three {\sc vnei} components
(since the abundances should be the same for all components
if they have a common origin),
and we set the abundances of elements from C to Ni to vary.
Different $kT_{\rm{e}}$ and $n_{\rm{e}}t_{\rm{ion}}$ were assumed
for these components.
Although model (c) improved the fit,
it was still unacceptable.
The $\chi^2/d.o.f.$ value for this model is summarized
in Table~\ref{table:chi} under (c) 3ISM or 3Ejecta.
For the three components in this model, the best-fit values for the temperature $kT_{\rm{e}}$  were
estimated to be
roughly $\sim$0.22, $\sim$0.67, and $\sim$2.75~keV, and the ionization parameter $\log(n_{\rm{e}}t~[\rm{cm^{-3}\,s}])$ was estimated to be
 $\sim$13, $\sim$13, and $\sim$11.
In this regard, based on XMM-Newton EPIC MOS and RGS observations, there have been already reports on
the temperature ($0.55^{+0.05}_{-0.32}$, $0.65\pm0.05$, and $3.5\pm0.5$)
and the ionization parameter ($10.36^{+1.64}_{-0.06}$, $>12.34$,
and $10.72\pm0.06$) \citep{2002A&A...392..955V}.
The values obtained from XMM-Newton observations roughly agree with
the values obtained with model (c).
According to \citet{2002A&A...392..955V},
the abundance of Fe is distributed among two cool components
and one hot component.
We therefore decoupled the abundance of Fe and performed another fit,
but this 
also failed to reproduce the spectra
($\chi^2/d.o.f. = 1257.519/464~(= 2.71)$).
This discrepancy could be due to higher statistics
but lower spectral resolution
of our study
since a grating spectrometer was used in the XMM-Newton observations. 

With models (d) and (e),
in order to reduce the degree of freedom of the model parameters,
the heavy element abundances were assumed to be the same for
the two {\sc vnei} components of the same type (i.e., 2ISM and 2Ejecta).
In the case of one ISM and two ejecta components,
we were unable to obtain reasonable constraints for the abundances of He, C, N, S, Ca, and Ni through spectral analysis.
We therefore fixed these parameters of the ISM component
to the average values for LMC
(He = 0.89, C = 0.30, N = 0.12, S = 0.31, Ca = 0.34,
and Ni = 0.62~times solar).
Furthermore, in the case of one ejecta and two ISM components,
reasonable constraints could not be obtained for the abundances of He, C, N, and O of the ejecta component,
and thus we fixed each of these abundances at 1~solar.
Neither model was able to provide an acceptable fit,
although the $\chi^2/d.o.f.$ values were significantly improved.
The $\chi^2/d.o.f.$ values for these models are summarized
in Table~\ref{table:chi} under (d) 1ISM + 2Ejecta and (e) 2ISM + 1Ejecta.

\subsubsection{Models with Four and Five Plasma Components
}\label{sec:fts}

In the previous section,
it was demonstrated that a reasonable fit of the emission requires yet
another component.
We therefore adopted a fourth thermal plasma component.
Since model (e) 2ISM + 1Ejecta provided the best fit,
we added another {\sc vnei} component
(ejecta in model (f) and ISM in model (g))
to model (e).
The heavy element abundances were taken to be the same for
both ISM and ejecta,
and the free parameters were the same as those in the models presented in \S\ref{sec:tts}.
Model (f)
resulted in only slight improvement compared with model (e).
In contrast, model (g) resulted in a
significant improvement in the closeness of fitting.
The $\chi^2/d.o.f.$ values for these models are
summarized in Table~\ref{table:chi} under (f) 2ISM + 2Ejecta and
(g) 3ISM + 1Ejecta.

Finally, we added another {\sc vnei} component
 (ejecta in model (h) and ISM in model (i)) to model (g).
However, in either case there was no significant improvement in fitting
compared to model (g).
The $\chi^2/d.o.f.$ values for these models are summarized
in Table~\ref{table:chi}
under (h) 3ISM + 2Ejecta and (i) 4ISM + 1Ejecta.

We regarded model (g) as the best-fit spectral model.
There are still residuals in the low energy end of the Fe-K line,
which could not be closely reproduced
by any of the multi-temperature {\sc vnei} plasma models.
Thus, we added a {\sc Gaussian} in the {\sc xspec} library.

We present the best-fit spectra and parameters
in Figure~\ref{fig:n103b_spectra} (a) and Table~\ref{table:model1},
respectively.
The three ISM components are shown in green, blue, and orange, and
the ejecta component is shown in cyan.
The Gaussians (shown in magenta) represent the line emissions
from Ar K$\alpha$ and Fe-K.
The difference of best-fit values of $n_{\rm{H}}^{\rm{LMC}}$
and the normalization bewteen FI and BI data were
$\sim0.4\times10^{21}~\rm{H\,cm^{2}}$ and $\sim$1\%,
and the best-fit gain offset values of FI and BI were
$\pm$0.0~eV and $-$8.0~eV,
respectively.
These values are within the calibration uncertainty.
Note that the best-fit values did not change significantly
upon fixing the energy gain to 0
and taking the same normalization and $n_{\rm{H}}$  for FI and BI.

\subsubsection{Model with Four Plasma Components with
Pure Metal Plasma Assumption}\label{sec:pure_metal}

The spectral fitting in \S\ref{sec:fts} requires
a single ejecta component with abundances of $\sim$0.06--30~solar.
The intensity of thermal bremsstrahlung (free-free emission)
is proportional to
$\sum_{\rm{i = H}}^{\rm{i = Ni}} n_{\rm{e}}n_{\rm{i}}Z_{\rm{i}}^2$,
where i, $n_{\rm{i}}$, and $Z_{\rm{i}}$ represent an element,
its number density,
and the ion charge of element i, respectively.
The species denoted by i are H, He, C, N, O, Ne, Mg, Si, S, Ca, Fe, and Ni
in the {\sc vnei} plasma code version 1.1 in the {\sc xspec} library.
With the aid of the observed abundances ($A_{\rm{i}}$),
the solar abundances ($n_{\rm{i}\odot}$) and the hydrogen number density
($n_{\rm{H}}$),
$\sum_{\rm{i = H}}^{\rm{i = Ni}} n_{\rm{e}}n_{\rm{i}}Z_{\rm{i}}^2$ is
$\sum_{\rm{i = H}}^{\rm{i = Ni}} A_{\rm{i}}A_{\rm{i}\odot}Z_{\rm{i}}^2n_{\rm{e}}n_{\rm{H}}$.
If the plasma consists only of hydrogen,
$\sum_{\rm{i = H}}^{\rm{i = Ni}} A_{\rm{i}}A_{\rm{i}\odot}Z_{\rm{i}}^2n_{\rm{e}}n_{\rm{H}}$ is $1\times n_{\rm{e}} n_{\rm{H}}$.
The plasma therefore consists mainly of hydrogen
when the coefficient $A_{\rm{i}}A_{\rm{i}\odot}Z_{\rm{i}}^2$ of
$n_{\rm{e}} n_{\rm{H}}$ for elements from C to Ni is smaller than 1.
Thus, the plasma assumed in \S\ref{sec:fts} contains 
predominantly hydrogen (hence H-dominated plasma)
since $A_{\rm{i}} A_{\rm{i}\odot} Z_{\rm{i}}^2 < 0.5 \ll 1$ for elements from C to Ni.
In contrast,
when the plasma consists mainly of heavy elements,
such as in the case of ejecta from the inner core of the progenitor star,
$A_{\rm{i}}A_{\rm{i}\odot}Z_{\rm{i}}^2$ for elements from C to Ni is considerably greater than 1,
and we must apply special adjustments to the plasma model
in performing fitting with the {\sc xspec} package.
Here, we assumed that the plasma consisted of heavy elements from C to Ni
with $A_{\rm{i}}A_{\rm{i}\odot}Z_{\rm{i}}^2 \gg 1$
(pure metal plasma)
and examined which plasma model is more suitable in the case of N103B.
Pure metal plasma has also been discussed
in \citet{1996A&A...307L..41V}, \citet{2008PASJ...60S.141Y},
and \citet{2010A&A...519A..11K}.
In these studies,
they derived that the abundance of heavy elements are 
in the order of $10^4$.

In the previous section,
the 3ISM + 1Ejecta model produced the best fit.
In this model, the single ejecta component clearly corresponds to  ejecta
with abundances of $\sim$0.06--30~solar.
We therefore fixed the abundances of elements from C to O for the ejecta
at $10^4$~solar in order to model plasma
poor in hydrogen and rich in heavy elements,
and fitted the spectra again.
This is the same method shown in \citet{2008PASJ...60S.141Y}
and \citet{1996A&A...307L..41V}.

The abundance of elements from C to O
can take any value much greater than 1 (i.e., $A_{\rm{i}}A_{\rm{i}\odot}Z^2 \gg 1$);
for instance,  the abundances of elements from C to O can be fixed
at $10^5$ or even $10^6$~solar.
Thus, the values of $A_{\rm{i}}A_{\rm{i}\odot}Z_{\rm{i}}^2$ for elements from C to O
satisfy the assumption of pure metal plasma.
The spectra reproduce a unique solution
to obtain convergence to a constant value of the product of the abundance
of elements from C to Ni
and the normalization ($A_{\rm{i}} \times Norm = const$).
Because of this, we fixed these abundances at $\sim10^5$--$10^6$~solar
to estimate the statistical error of the emission measure.
For these abundances,
we also fixed the emission measure at 2.74$\times10^{55}~\rm{cm^3}$.
The absolute values of these abundances
and the emission measure obtained by the spectral analysis do not play an important role.
This method is commonly used for reproducing pure-metal plasma in the {\sc xspec} package
\citep{2008PASJ...60S.141Y}.

\begin{figure*}
 \begin{center}
    \FigureFile(140mm,140mm)
	{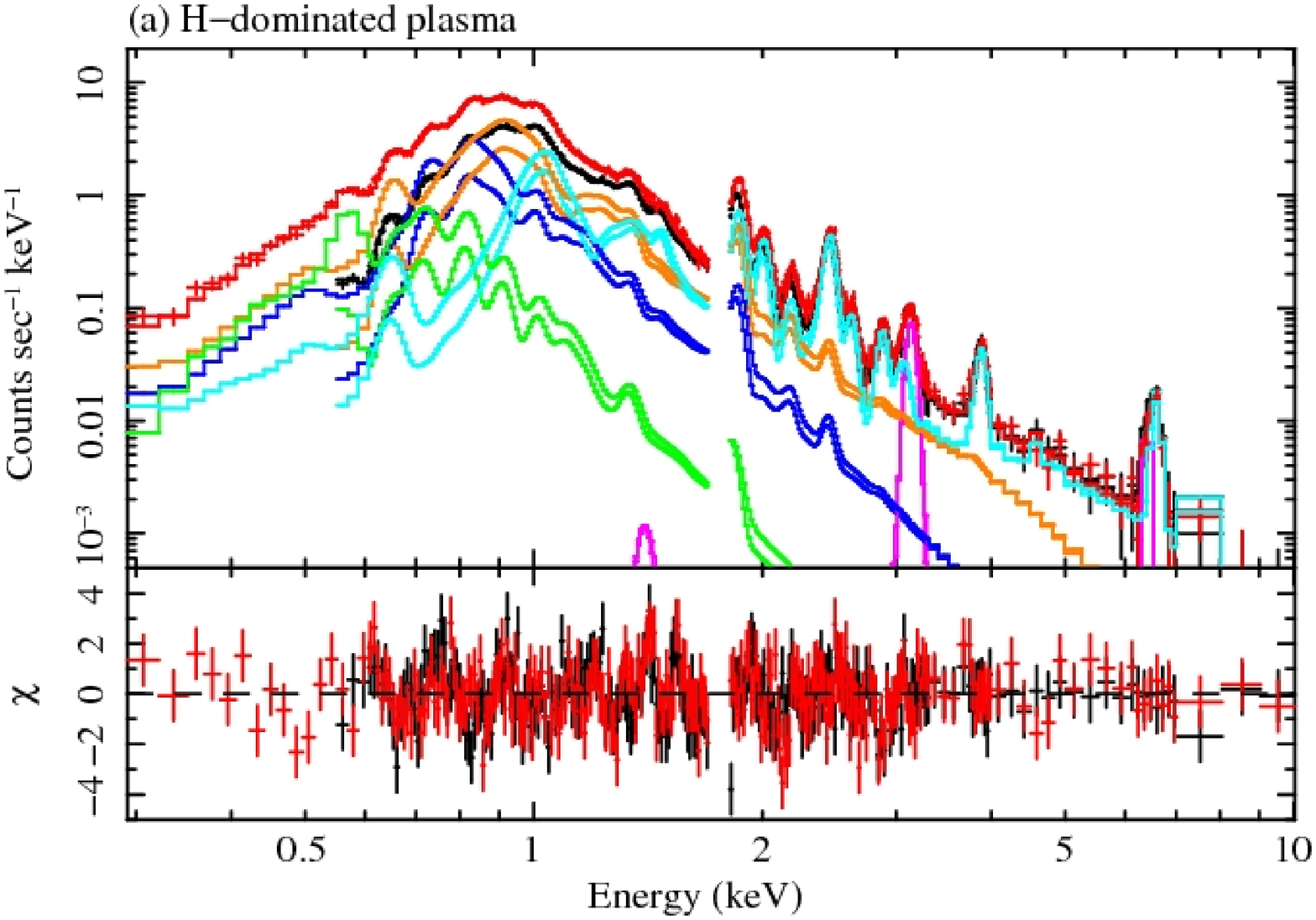}	
    \FigureFile(140mm,140mm)
	{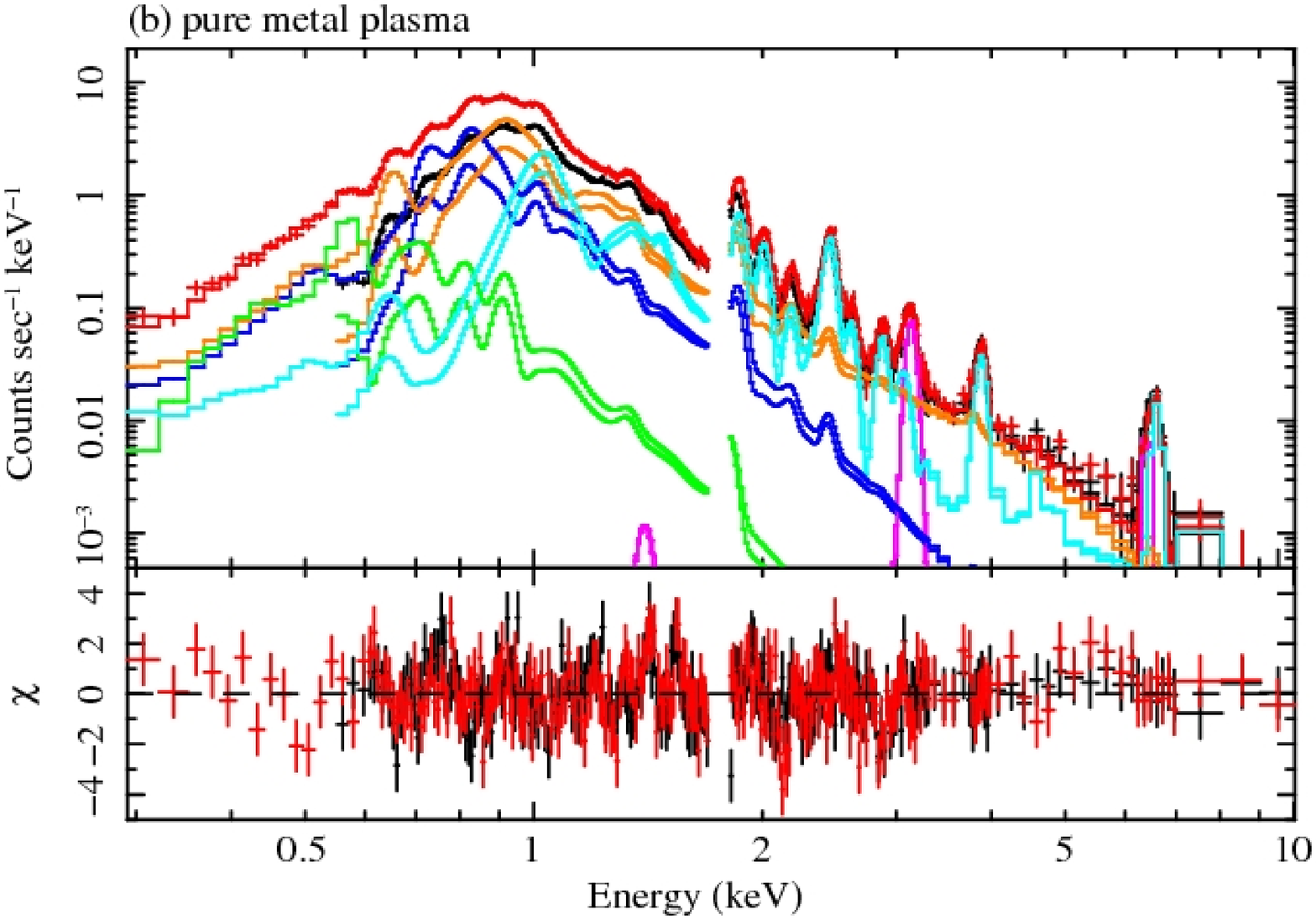}
  \end{center}
  \caption{(a) : H-dominated plasma model.
(b): Pure metal plasma model.
Upper panels: background-subtracted XIS spectra in the 0.3-10.0~keV band.
The black and red data points represent the FI and BI spectra, respectively.
The three ISM components are shown in green, blue and orange, and the
ejecta component is shown in cyan.
The Gaussians (shown in magenta) represent the lines for Ar K$\alpha$
and Fe K, respectively.
Lower panels: residuals from the best-fit model.
Data in red and black represent FI and BI,
respectively.}\label{fig:n103b_spectra}
\end{figure*}

Table~\ref{table:pure_metal} and Figure~\ref{fig:n103b_spectra} (b)
show the best-fit parameters and spectra, respectively.
The fitting shows a reduced $\chi^2$ similar to that in 
the H-dominant plasma model.
Note that
whereas the high-energy continuum in the 4.0--6.0~keV band corresponds
to ejecta in the H-dominant plasma model,
it corresponds to the high-temperature ISM component 
in the pure metal plasma model,
and the Fe-K line is from an ejecta component in both models.

\begin{table*}
  \caption{Best-fit parameters for N103B modeled
with H-dominated plasma$^*$. }\label{table:model1}
  \begin{center}
    \begin{tabular}{lcccc}
    \hline
         Parameter & \multicolumn{4}{c}{{\sc Absorption components}} \\
    \hline
	$N_{\rm{H}}^{\rm{Galactic}}$ \,($\times10^{20}~\rm{cm^{-2}}$) & \multicolumn{4}{c}{6.2 (fixed)} \\
	$N_{\rm{H}}^{\rm{LMC (FI)}}$ ($\times10^{21}~\rm{cm^{-2}}$) & \multicolumn{4}{c}{2.97$\pm0.06$}                 \\
	$N_{\rm{H}}^{\rm{LMC (BI)}}$ ($\times10^{21}~\rm{cm^{-2}}$) & \multicolumn{4}{c}{2.55$^{+0.04}_{-0.03}$}  \\
    \hline
		             & \multicolumn{4}{c}{{\sc Plasma components}} \\
	Parameter & \multicolumn{3}{c}{ISM components} & Ejecta component \\\hline
	$kT_{\rm{e}}$ (keV) & 0.319$_{-0.005}^{+0.007}$ & 0.547$^{+0.005}_{-0.006}$ & 0.962$\pm0.005$ & 3.96$\pm0.01$  \\
	$\log(n_{\rm{e}}t)$  ($\rm{cm^{-3}}s$) & 10.48$^{+0.05}_{-0.08}$ & $>$11.88 & 10.89$\pm0.01$ & 	10.795$^{+0.007}_{-0.006}$ \\
	$EM^\dagger$  &  1.17$^{+0.97}_{-0.06}$ & 2.33$^{+0.04}_{-0.05}$ & 2.60$\pm0.04$ & 0.197$^{+0.003}_{-0.001}$ \\
	H$^\ddagger$   & & 1.00 (fixed) &  & 1.00 (fixed) \\
	He & & 0.89 (fixed) & & 1.00 (fixed) \\
	C   & & 0.30 (fixed) & & 1.00 (fixed) \\
	N   & & 0.12 (fixed) & & 1.00 (fixed) \\
	O   &  & 0.153$^{+0.006}_{-0.005}$ & & 1.00 (fixed) \\
	Ne & & 0.180$^{+0.002}_{-0.004}$ & &  $<$0.06 \\
	Mg & & 0.134$^{+0.011}_{-0.010}$ & & 2.27$\pm0.17$  \\
	Si   & & 0.63$^{+0.03}_{-0.02}$        &  &9.7$\pm0.3$  \\
	S    & & 0.31 (fixed) &  &17.2$^{+0.5}_{-0.4}$\\
	Ca & & 0.34 (fixed) &  & 30$\pm3$\\
	Fe  & & 0.523$^{+0.005}_{-0.006}$ &  & 4.19$^{+0.07}_{-0.09}$\\
	Ni   & & 0.62 (fixed) &  & 18.6$^{+1.8}_{-1.6}$\\
    \hline	
	Line & \multicolumn{3}{c}{Center energy (keV)} & Normalization (photons\,cm$^{-1}$\,s$^{-1}$) \\
	Ar He-like K$\alpha$ & \multicolumn{3}{c}{3.141$\pm0.005$} & 3.8$^{+0.3}_{-0.4}\times10^{-5}$\\
	Fe-K & \multicolumn{3}{c}{6.43$^{+0.02}_{-0.03}$} & 6.8$^{+1.7}_{-2.2}\times10^{-6}$\\
    \hline
	Const (FI) & \multicolumn{4}{c}{1.00 (fixed)} \\
	Const (BI) & \multicolumn{4}{c}{1.000$^{+0.007}_{-0.004}$} \\
    \hline	
	FI gain offset (eV) & \multicolumn{4}{c}{$-7.475\pm0.005$} \\
	BI gain offset (eV) & \multicolumn{4}{c}{$-0.2^{+0.2}_{-0.8}$} \\
    \hline	
         $\chi^2 / d.o.f. $	& \multicolumn{4}{c}{602.272 / 458 (= 1.32)} \\
      \hline
    \end{tabular}
  \end{center}
  {$^*$ Errors represent 90\% confidence intervals.}\\
  {$^\dagger$ Emission measure
$EM = \int n_{\rm{e}} n_{\rm{H}} dV = n_{\rm{e}} n_{\rm{H}} V$
in units of 10$^{59}$~cm$^{-3}$.
The distance to N103B was assumed to be 48~kpc \citep{2006ApJ...652.1133M}.}\\
  {$^\ddagger$ Abundance is relative to the solar abundance
\citep{1989GeCoA..53..197A},
and a common value is assumed for all three ISM components.}
\end{table*}

\begin{table*}
  \caption{Best-fit parameters for N103B modeled with
pure metal plasma$^*$. }\label{table:pure_metal}
  \begin{center}
    \begin{tabular}{lccccc}
    \hline
    Parameter & \multicolumn{4}{c}{{\sc Absorption components}} \\
    \hline
	$N_{\rm{H}}^{\rm{Galactic}}$ ($\times10^{20}~\rm{cm^{-2}}$)          & \multicolumn{4}{c}{6.2 (fixed)} \\
	$N_{\rm{H}}^{\rm{LMC (FI)}}$ ($\times10^{21}~\rm{cm^{-2}}$) & \multicolumn{4}{c}{3.07$^{+0.04}_{-0.10}$} \\
	$N_{\rm{H}}^{\rm{LMC (BI)}}$ ($\times10^{21}~\rm{cm^{-2}}$) & \multicolumn{4}{c}{2.74$^{+0.05}_{-0.11}$} \\
    \hline
	& \multicolumn{4}{c}{{\sc Plasma components}}\\
	Parameter & \multicolumn{3}{c}{ISM components} & Ejecta component\\
    \hline
	$KT_{\rm{e}}$ (keV) & 0.328$^{+0.001}_{-0.013}$ & 0.522$^{+0.01}_{-0.001}$ & 1.29$^{+0.01}_{-0.02}$ & 4.01$^{+0.05}_{-0.08}$ \\
	$\log(n_{\rm{e}}t)$ ($\rm{cm^{-3}}s$) & 10.11$^{+0.08}_{-0.09}$ & $>$11.86 & 10.654$^{+0.006}_{-0.005}$ & 10.785$^{+0.005}_{-0.008}$ \\
	$EM^\dagger$     & 8.55$^{+0.66}_{-1.10}$ & 23.9$^{+0.9}_{-1.1}$ & 17.3$\pm 0.7$ & 2.74$^{+0.11}_{-0.04}\times10^{-5}$ \\
	H$^\ddagger$   & & 1.00 (fixed) & & 1.00 (fixed) \\
	He & & 0.89 (fixed) & & 1.00 (fixed) \\
	C   & & 0.30 (fixed) & & 1.00$\times10^{4}$ (fixed) \\
	N   & & 0.12 (fixed) & & 1.00$\times10^{4}$ (fixed) \\
	O   & & 0.131$^{+0.005}_{-0.006}$ & & 1.00$\times10^{4}$ (fixed) \\
	Ne & & 0.18$^{+0.01}_{-0.02}$ & & $< 2.6\times10^{4}$\\
	Mg & & 0.11$\pm0.1$& & 1.3$^{+0.3}_{-0.1}\times10^{5}$ \\
	Si   & & 0.55$^{+0.03}_{-0.02}$ & & 5.6$^{+0.1}_{-0.2}\times10^{5}$ \\
	S    & & 0.31 (fixed) & & 9.9$\pm0.3\times10^{5}$\\
	Ca & & 0.34 (fixed) & & 1.7$\pm0.1\times10^{6}$\\
	Fe  & & 0.512$^{+0.005}_{-0.004}$ & & 2.46$^{+0.06}_{-0.05}\times10^{5}$\\
	Ni   & & 0.62 (fixed) &  & 8.6$^{+1.3}_{-0.6}\times10^{5}$\\
    \hline
	Line & \multicolumn{3}{c}{Center energy (keV)} & Normalization (photons\,cm$^{-1}$\,s$^{-1}$) \\
	Ar He-like K$\alpha$ & \multicolumn{3}{c}{3.141$\pm0.007$} & (3.7$\pm0.5)\times10^{-5}$\\
	Fe-K & \multicolumn{3}{c}{6.43$\pm0.04$} & 7.0$\pm3.0\times10^{-6}$\\
    \hline
	Const (FI) & \multicolumn{4}{c}{1.00 (fixed)} \\
	Const (BI) & \multicolumn{4}{c}{1.008$\pm$0.01} \\
    \hline
	FI gain offset (eV) & \multicolumn{4}{c}{$-7.2\pm0.1$} \\
	BI gain offset (eV) & \multicolumn{4}{c}{$-0.5^{+0.4}_{-0.5}$} \\
    \hline
         $\chi^2 / d.o.f. $	& \multicolumn{4}{c}{644.936 / 458 (= 1.41)} \\
      \hline
    \end{tabular}
  \end{center}
  {$^*$ Errors represent 90\% confidence intervals.
In the estimation of the statistical error of the normalization,
we fixed the abundance at $\sim10^5$--10$^6$~solar.
In the estimation of the statistical error of the abundance,
we fixed the emission measure.}\\
  {$^\dagger$ Emission measure
$EM = \int n_{\rm{e}} n_{\rm{H}} dV = n_{\rm{e}} n_{\rm{H}} V$
in units of 10$^{57}$~cm$^{-3}$.
The distance to N103B was assumed to be 48~kpc \citep{2006ApJ...652.1133M}.}\\
    {$^\ddagger$ Abundance is relative to the solar abundance
\citep{1989GeCoA..53..197A},
and a common value is assumed for all three ISM components.}
\end{table*}

\subsection{Spectral Analysis Around Fe-K Line}
\label{sec:spec:iron}

In the previous section, we showed that
the emission was closely reproduced with four plasma components,
but still had residuals
at the low-energy end of the Fe-K line.
This implies that
the plasma emitting the Fe-K line has a relatively low degree of ionization.
We therefore fitted the spectrum in the 4.0--10.0~keV band
using a plane-parallel non-equilibrium ionization {\sc vpshock} model
in {\sc xspec},
which integrates the emission with ionization parameters varying linearly
from 0 to $n_{\rm{e}}t_{\rm{ion}}$ (see also \citet{2000ApJ...543L..61H}).

\begin{figure}
 \begin{center}
    \FigureFile(80mm,80mm)
    	{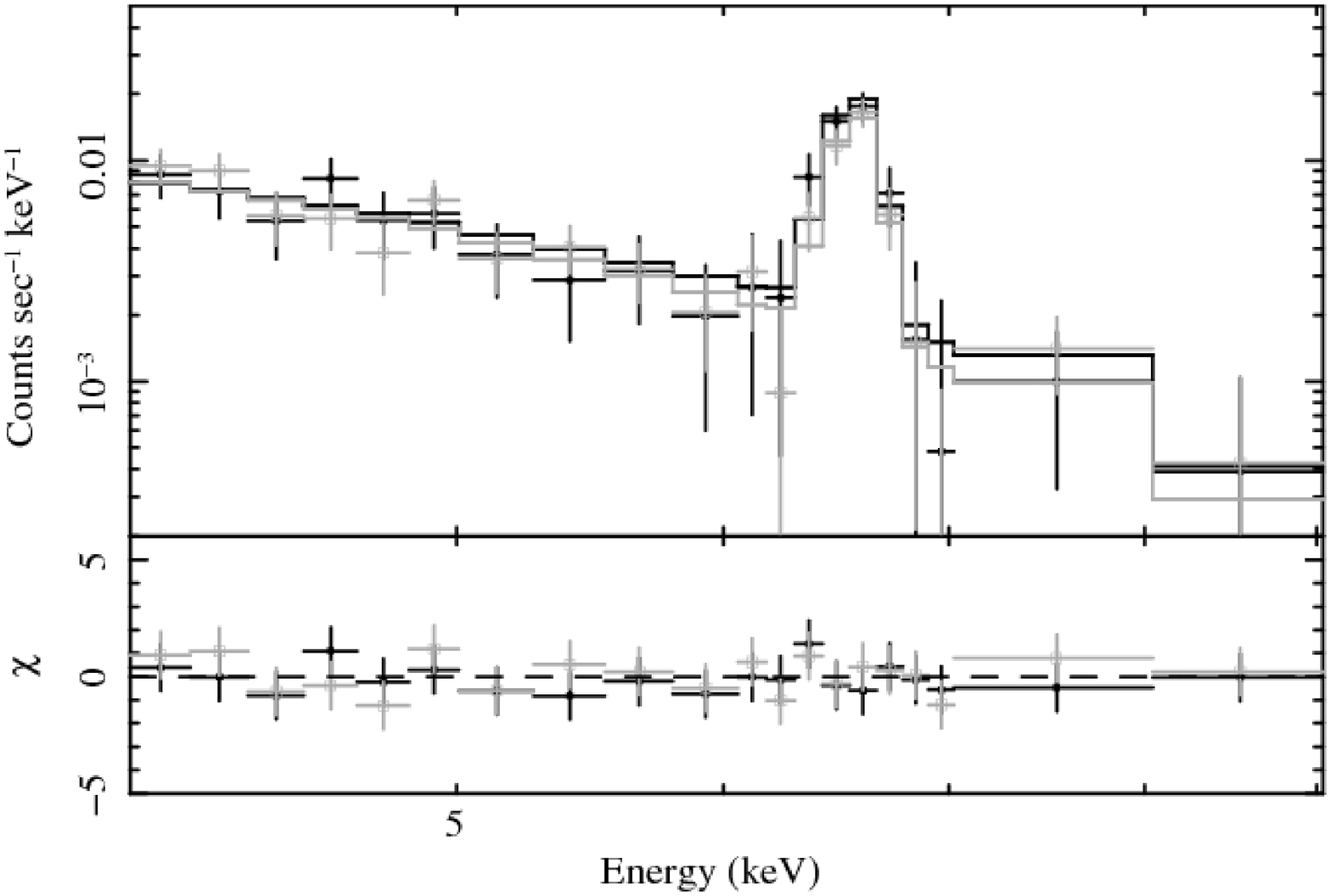}
  \end{center}
  \caption{Upper panel:
Background-subtracted XIS spectra in the 4.0--10.0~keV band.
Lower panel: Residuals from the best-fit models.
In both panels, data in black and light gray represent the FI and BI spectra, respectively.}
  \label{fig:propagation}
\end{figure}

In the spectral analysis,
we fixed the temperature at 
the best-fit value of 3.96~keV (see Table~\ref{table:model1}).
We fitted the spectrum
by taking the normalization and $n_{\rm{e}}t_{\rm{ion}}$ as variables.
As a result, the fit was accepted
with $\chi^2/d.o.f.$ of 18.46/39 (= 0.43).
The best-fit values of the ionization parameter,
the emission measure, and the abundance of Fe were estimated
to be 10.81 (10.73--10.89),
0.39 (0.35--0.43)$\times10^{59}~\rm{cm}^{-3}$,
and 4.74 (3.71--5.96)~solar, respectively.
These best-fit values were close to the values obtained
with the H-dominated plasma model.
Figure~\ref{fig:propagation} shows the best-fit spectra,
where the fit of the residual at the low-energy end of Fe-K
was clearly improved with the {\sc vpshock} model.

\section{Image Analysis Using Chandra/ACIS Data}\label{sec:img}

\begin{table*}
 \caption{Information necessary for preparing continuum-subtracted images.}
 \label{table:acis_imgana}
   \begin{center}
    \begin{tabular}{lccccc}
    	\\\hline
	\multicolumn{6}{c}{{\sc Continuum component$^*$}} \\\hline
	\multicolumn{6}{c}{Photon index = 3.7$\pm0.1$, Normalization = 5.4$^{+0.6}_{-0.5}$~($\times10^{-3}$~photons\,keV$^{-1}$\,cm$^{-2}$\,s$^{-1}$)} \\\hline
	\multicolumn{6}{c}{{\sc Information necessary for preparing continuum-subtracted images}} \\\hline
	Element & Line ($I_{\rm{L}}$) & Continuum ($I_{\rm{C}}$) & Flux ($F_{\rm{C}}^{\star}$) & Flux ($F_{\rm{C}}$) & $R = \frac{F_{\rm{C}}^{\star}}{F_{\rm{C}}}$ \\
	& (eV) & (eV) & ($\times10^{-5}~\rm{photons\,cm^{-2}\,s^{-1}}$)  &  ($\times10^{-5}~\rm{photons\,cm^{-2}\,s^{-1}}$) \\\hline
	Si & 1750--1915 & 1560--1735 & 9.19 & 14.60 & 0.63 \\
	S & 2340--2540 & 2625--2825 & 3.78 & 2.49   & 1.52\\
	Ar & 3025--3225 & 3350--3550 & 1.48 & 1.02   & 0.15 \\
	Ca& 3750--3950 & 3350--3550 & 0.68 & 1.02  & 0.66 \\
	Fe & 6200--6800 & 5400--6000 & 0.28 & 0.46  & 0.61 \\\hline
    \end{tabular}
  \end{center}
  {$^*$ Errors represent 90\% confidence intervals.}
\end{table*}

In this section, we analyzed images obtained with Chandra/ACIS
to reveal the distribution of heavy elements from O to Fe-K
heated by the forward and reverse shocks.
In the first step in studying the distribution,
we prepared continuum-subtracted images (CS images) for elements from Si to Fe-K
in the same way as in \citet{2000ApJ...537L.119H}
and \citet{2003ApJ...582..770L}.
Below, we briefly describe the procedure for preparing CS images.
More detailed explanations are presented
in \citet{2000ApJ...537L.119H} and \citet{2003ApJ...582..770L}. 

\begin{figure}
  \begin{center}
   \FigureFile(80mm,80mm)
   {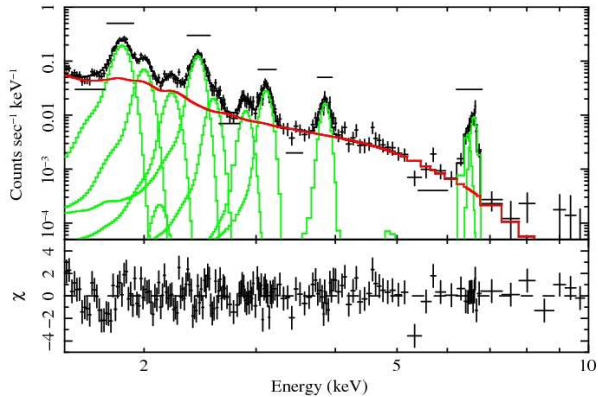}
  \end{center}
  \caption{Background-subtracted Chandra/ACIS spectrum of N103B.
The horizontal bars above and below the spectrum show
the energy interval for line and continuum images
used for preparing continuum-subtracted images (see text).}
 \label{fig:chandra_spectrum}
\end{figure}

First, we produced raw line energy band images ($I_{\rm{L}}$)
and continuum energy band images ($I_{\rm{C}}$) for each element.
The energy intervals for $I_{\rm{L}}$ and $I_{\rm{C}}$ are summarized in
the second and third columns in Table~\ref{table:acis_imgana}
and the horizontal bars above and below in the spectrum of
Figure~\ref{fig:chandra_spectrum}, respectively.
We selected the energy band of $I_{\rm{C}}$ to be as close as possible to that of $I_{\rm{L}}$,
to avoid line features.
The same continuum component as for Ca was also used for Ar
to avoid the line emission from Ca He$\beta$ ($\sim$4.58~keV).

In the next step,
using spectral analysis, we estimated the continuum flux ($F_{\rm{C}}^{\star}$)
included in $I_{\rm{L}}$.
We extracted the spectrum of the entire remnant
and obtained the response matrix using the {\sc specextract} command
in CIAO version 4.1.2.
The spectra were fitted with a {\sc power low} for the continuum emission
and {\sc Gaussian}s for line emissions in the {\sc xspec} library
in the 1.5--10.0~keV band.
Figure~\ref{fig:chandra_spectrum} shows the best-fit background-subtracted
spectrum.
The best-fit value for the photon index and the normalizations are summarized
in Table~\ref{table:acis_imgana}.
The fluxes included in $I_{\rm{L}}$ ($ = F_{\rm{C}}^{\star}$)
and $I_{\rm{C}}$ ($ = F_{\rm{C}}$) are summarized
in the fourth and fifth columns in Table~\ref{table:acis_imgana}. 

\begin{figure*}
 \begin{center}
     \FigureFile(55mm,55mm)
    	{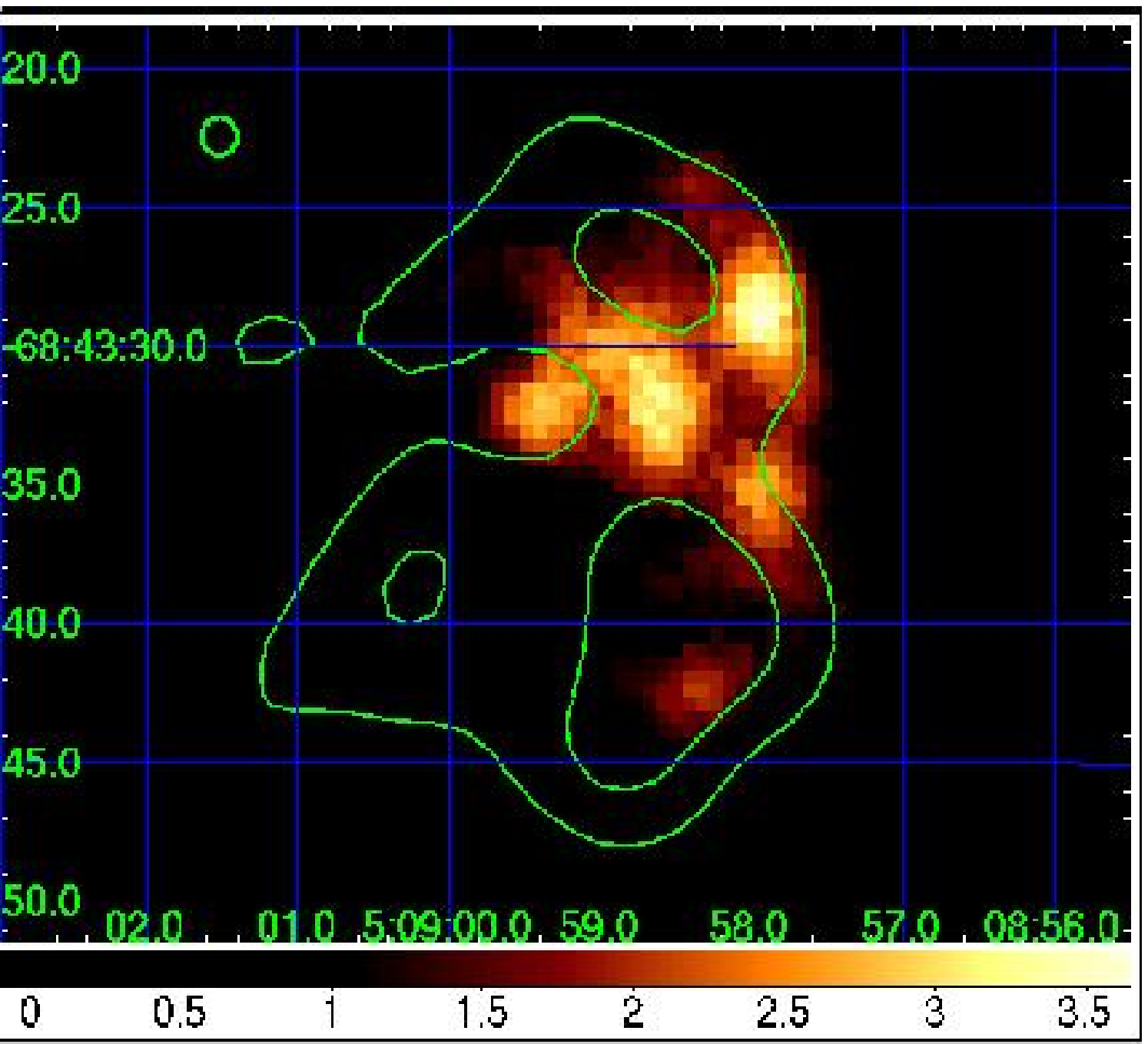}
    \FigureFile(55mm,55mm)
    	{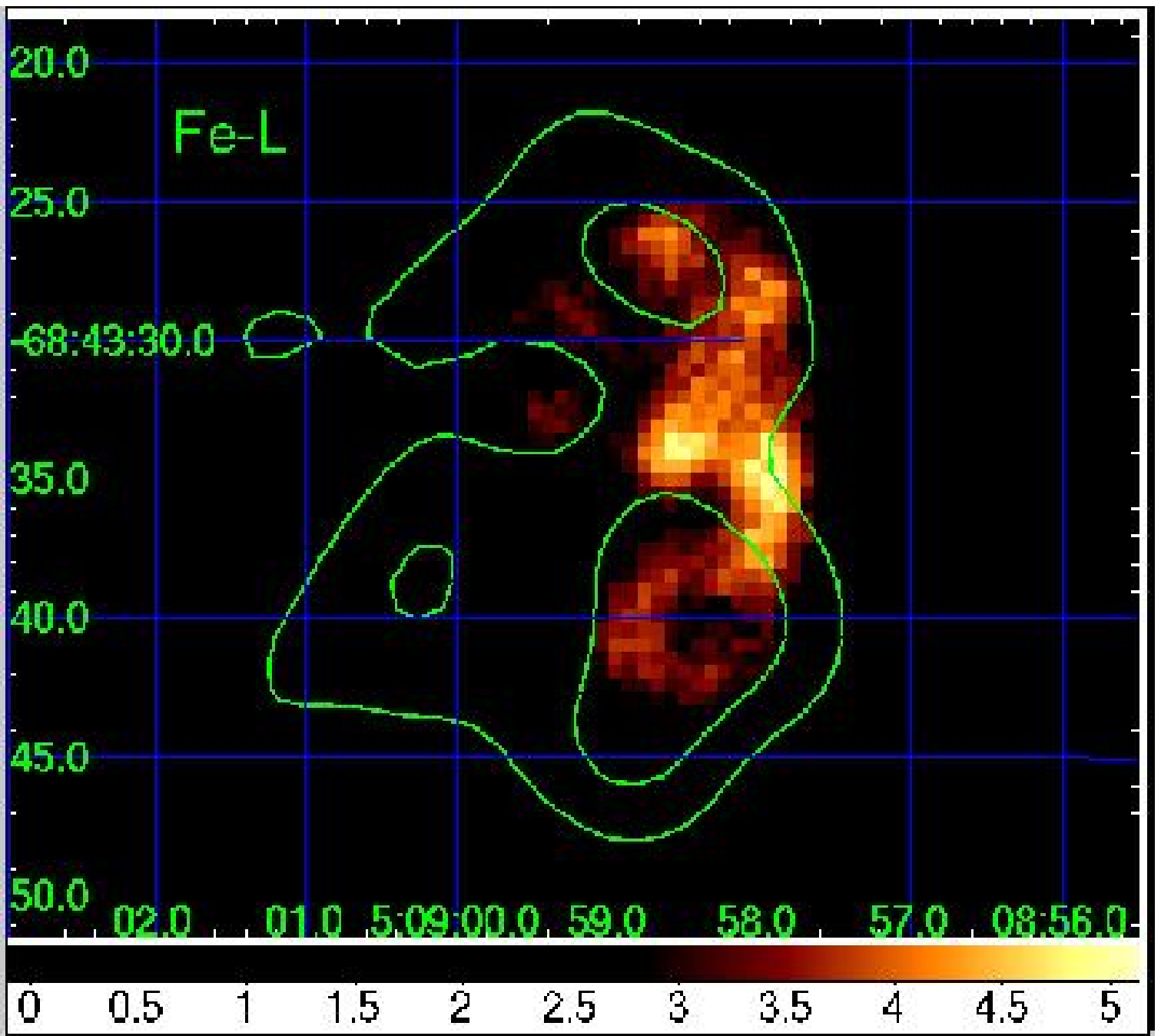}	
    \FigureFile(55mm,55mm)
    	{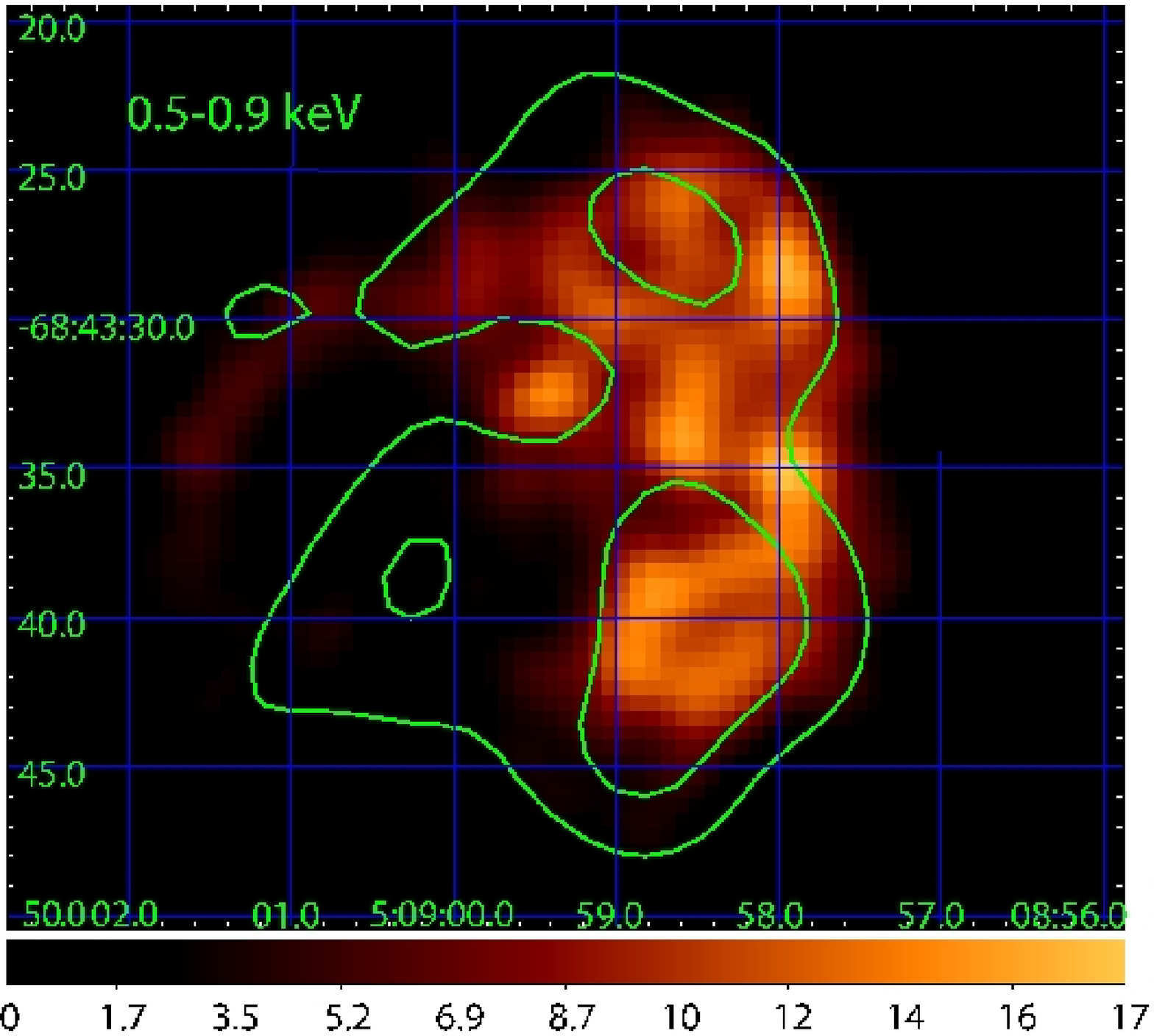}	
	\\
    \FigureFile(55mm,55mm)
    	{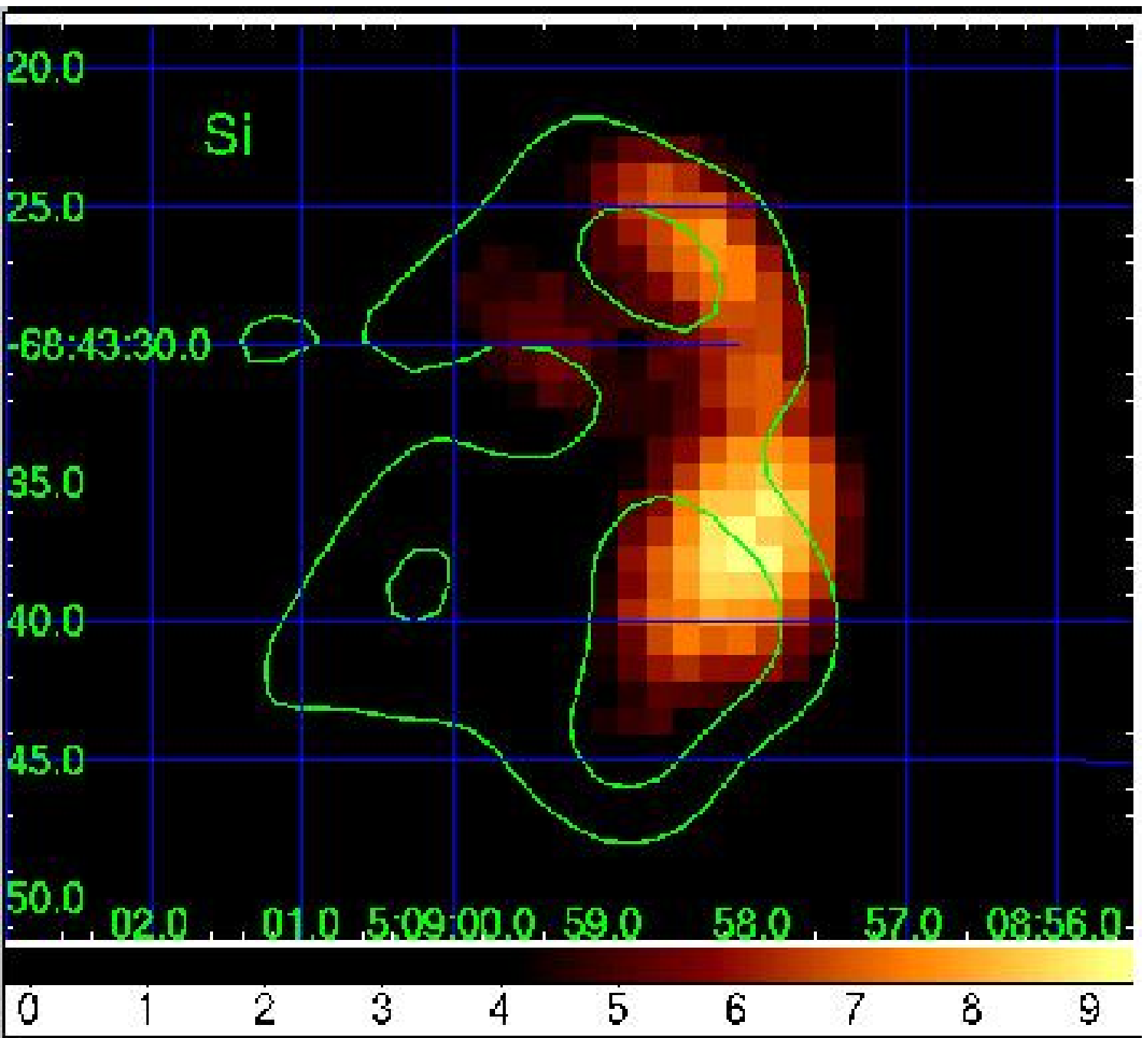}
    \FigureFile(55mm,55mm)
    	{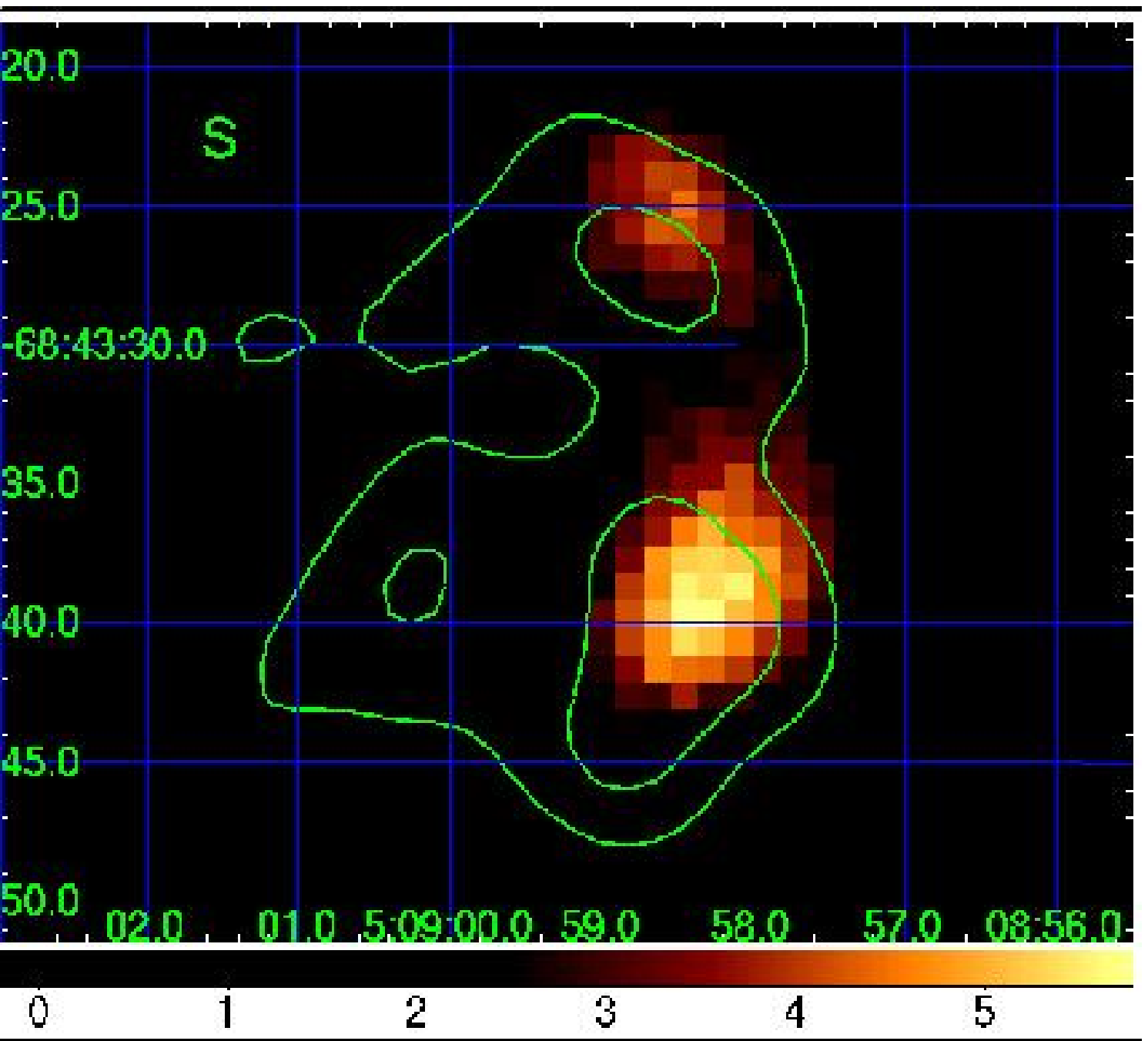}
    \FigureFile(55mm,55mm)
    	{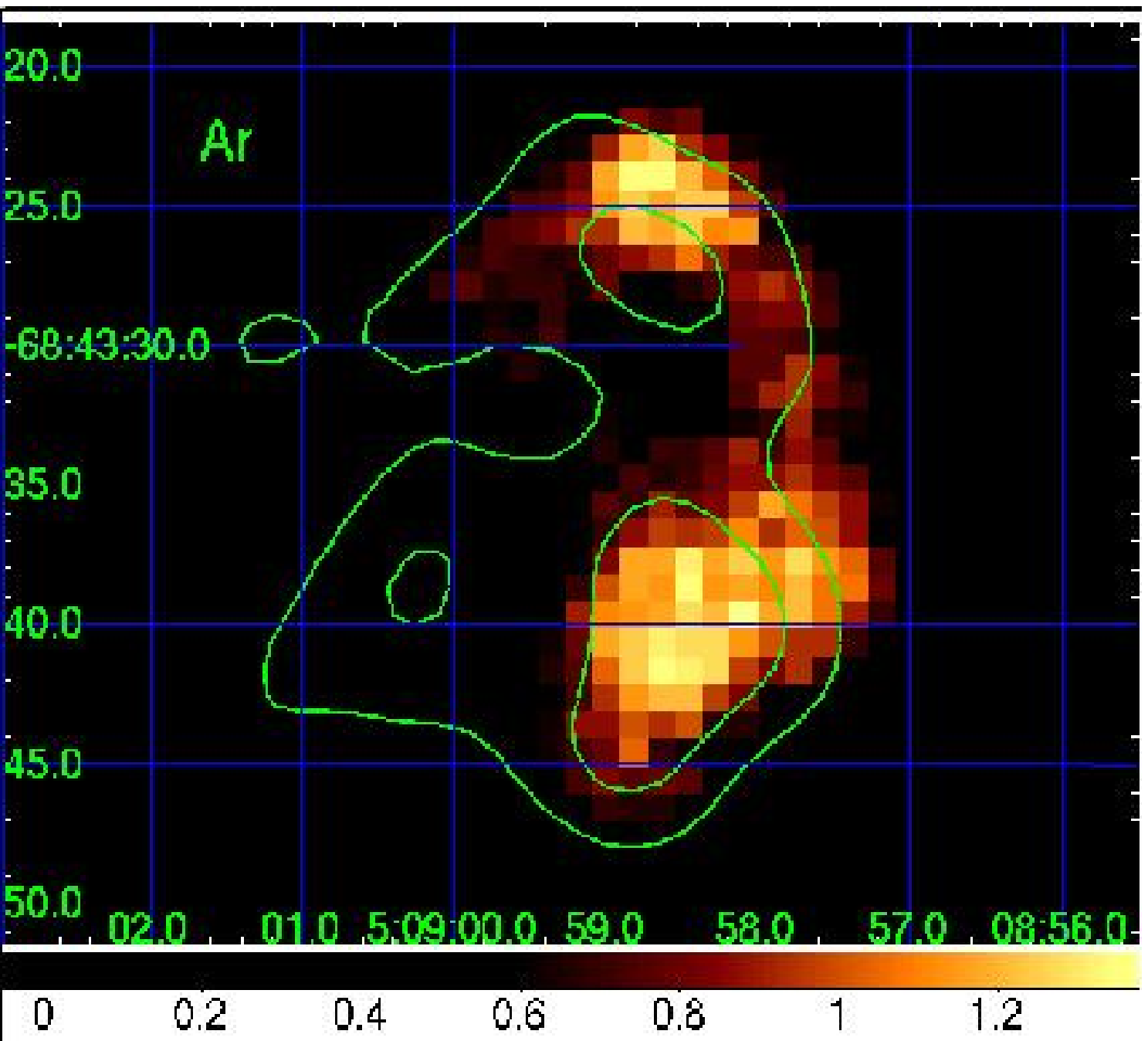}
    \FigureFile(55mm,55mm)
    	{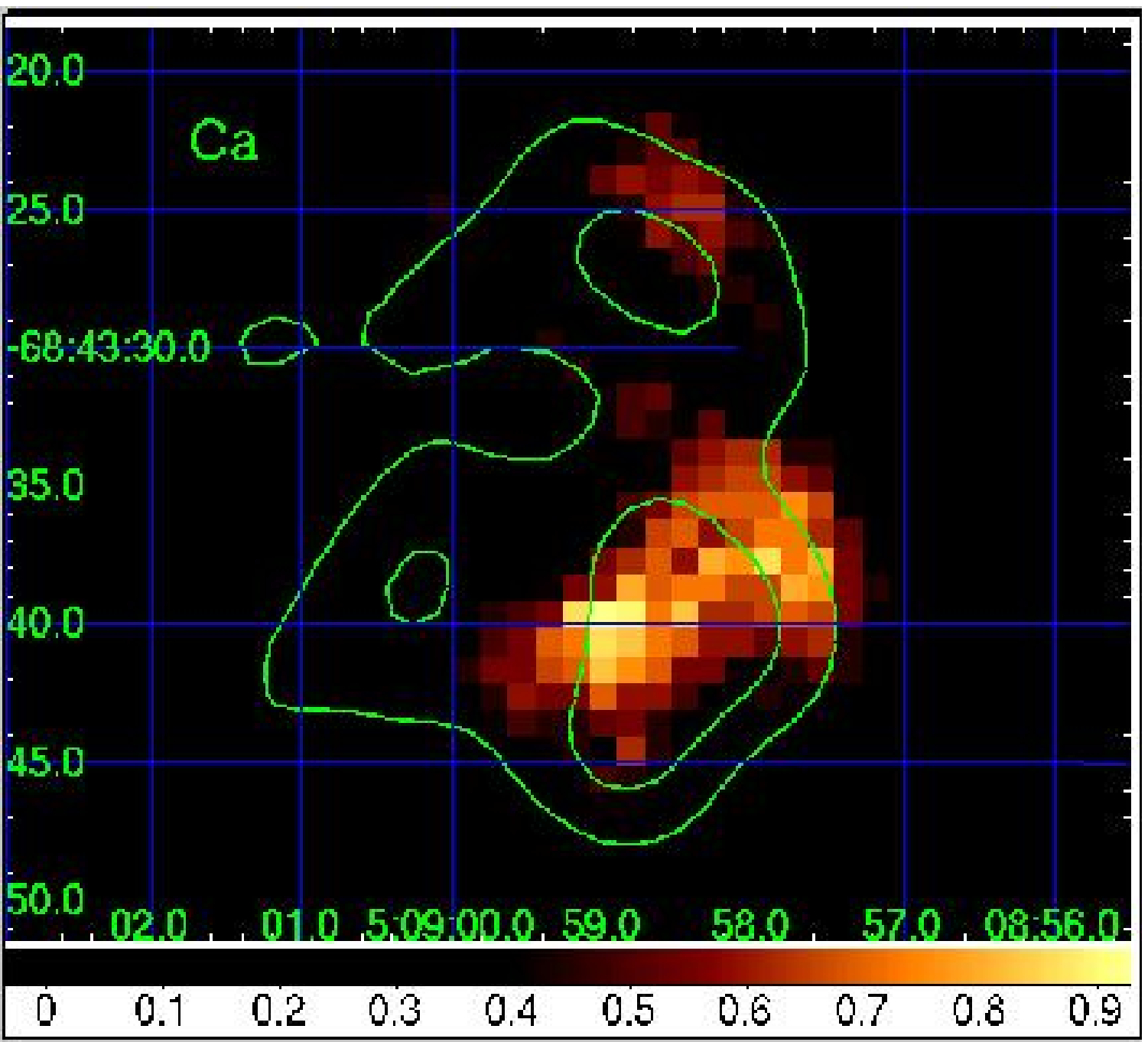}
    \FigureFile(55mm,55mm)
    	{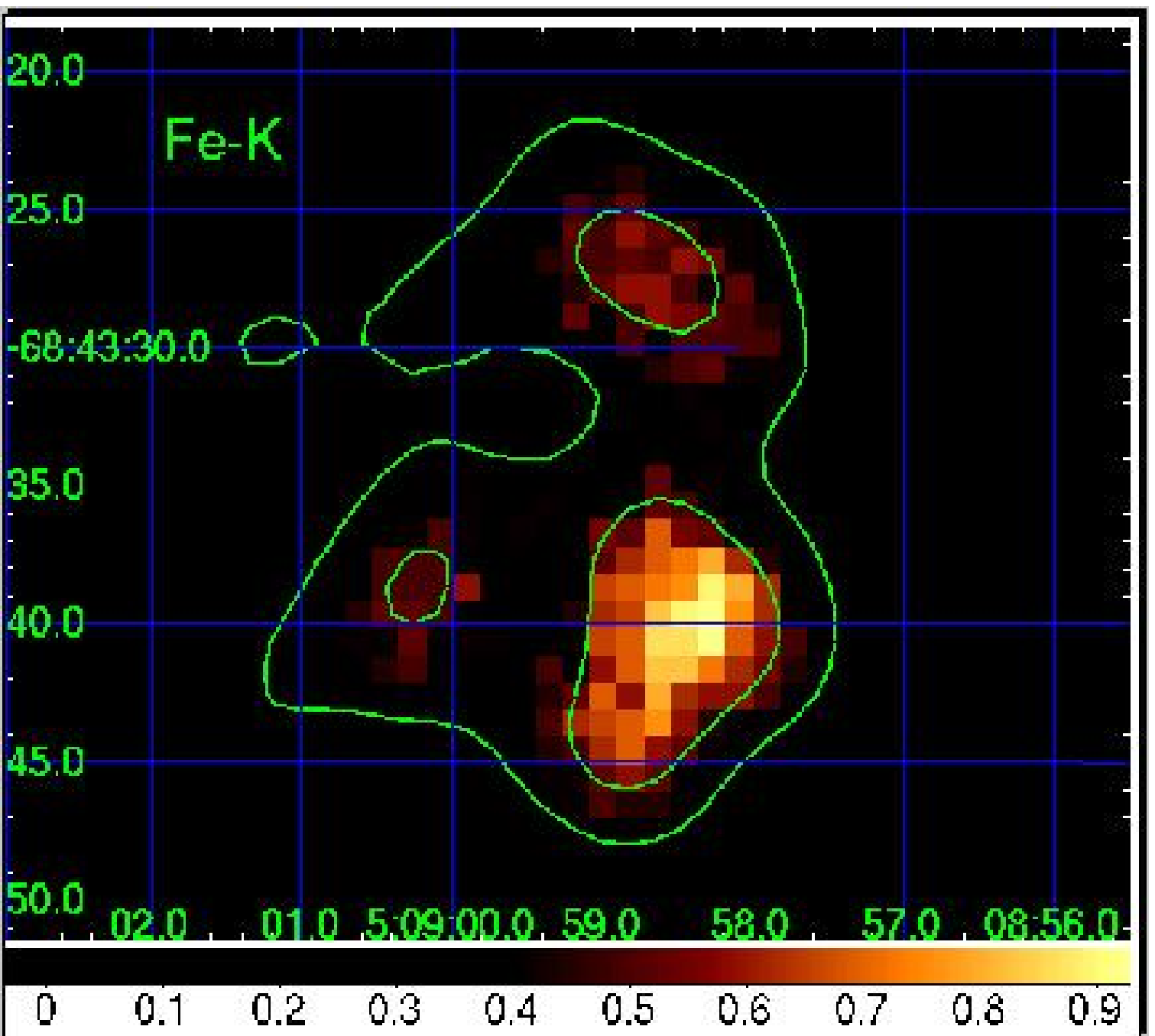}	
     \FigureFile(55mm,55mm)
   	{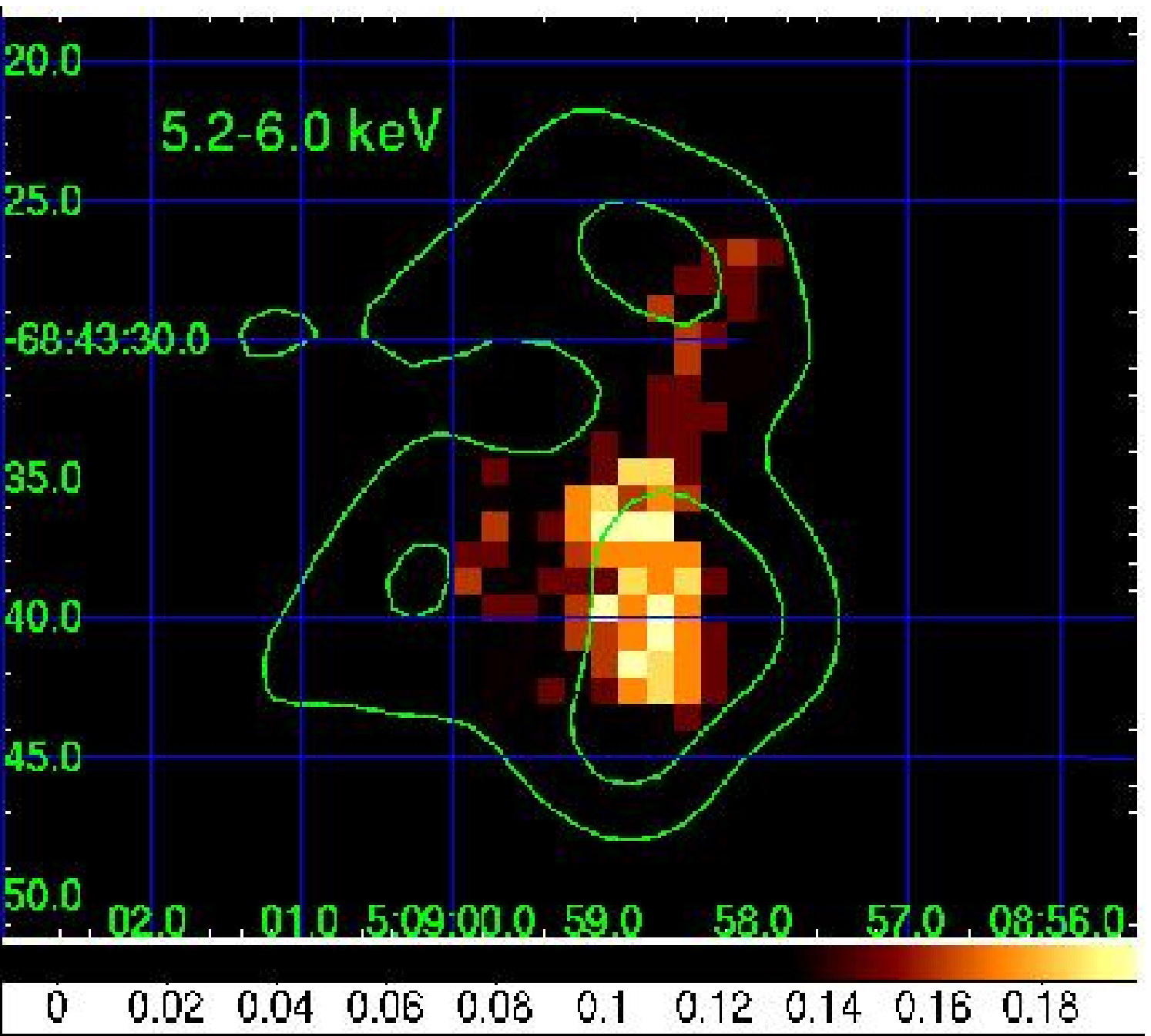}		
  \end{center}
  \caption{Chandra/ACIS narrow-band images of N103B.
The images labeled O and Fe-L are for O (0.5--0.7~keV)
and Fe-L complex (0.7--0.8~keV) band images.
These images are not smoothed.
The image labeled 0.5--0.9~keV corresponds to
the low-energy band image in the 0.5--0.9~keV.
The images labeled Si, S, Ar, Ca, and Fe-K are
continuum-subtracted images.
Each image is smoothed using a top-hat function with $\sigma = 3$,
and a bin size of 1''.
The image labeled 5.2--6.0~keV corresponds to the high-energy continuum
in the 5.2--6.0~keV band.
This image is smoothed using a top-hat function with $\sigma = 5$,
and a bin size of 2.5''.
The overlapping contours represent
0.2 and 0.4~counts\,px$^{-1}$ of the Fe-K line.
The coordinates are in the equatorial coordinate system for epoch 2000.}
  \label{fig:CSimage}
\end{figure*}

A continuum image ($I_{\rm{C}}^{\star}$)
included in the energy band of $I_{\rm{L}}$ can be obtained
from $R \times I_{\rm{C}}$,
where $R$ is $\frac{F_{\rm{C}}^{\star}}{F_{\rm{C}}}$.
Thus, the pure line image is given by $I_{\rm{L}} - I_{\rm{C}}^{\star}$.
Figure~\ref{fig:CSimage} shows the pure line images for Si, S, Ar, Ca, and Fe-K.

Regarding the pure line images for O and Fe-L,
we were unable to prepare CS images
since it was rather difficult to estimate $I_{\rm{C}}$ for O and Fe-L
with the Chandra spectral resolution.
We therefore prepared only energy-selected images at 0.5--0.7 keV for O
and 0.7--0.8 keV for Fe-L
(Figure~\ref{fig:CSimage}).
We also prepared low-energy band image in the 0.5--0.9~keV
and high-energy continuum image in the 5.2--6.0~keV band
(Figure~\ref{fig:CSimage}).

\section{Discussion}\label{sec:dis}

\subsection{Complex Emission Features}\label{sec:plasma_origin}

In this section,
we discuss the complex emission features
based on spectral and image analysis.

\subsubsection{Emission Origin}\label{sec:abs_abundance}

In \S~\ref{sec:spe},
we found that the plasma emission from N103B can be closely reproduced
by three ISM components and one ejecta component.
In this section,
we discuss the plasma origin.

The elemental abundances reflect the plasma in the environment
and the progenitor of the SNR.
In Tables~\ref{table:model1} and \ref{table:pure_metal},
the abundances for the three ISM components are significantly lower than
the solar abundances, and roughly consistent with
the LMC average up to a factor of 2,
whereas the abundances for the ejecta component are
notably higher than the solar abundances.
These results provide clues regarding the emission origin,
suggesting that the three ISM and one ejecta components are
emitted from ISM heated by the forward shock and ejecta heated
by the reverse shock, respectively.
Each ISM plasma component has different values of $kT$ and $n_e t$.
This might be due to the different time scales
of heating by the forward shock.
The middle-temperature component has a large $n_e t$,
which suggests that it may have been heated immediately
after the supernova explosion
or represents the emission of the dense gas in the ISM.
In contrast, the hot and cool components have
smaller $n_e t$.
Their respective emission measures are similar to
that of the middle-temperature component,
and thus it is not due to a difference
in density but a difference in time scale.
These two components may have been heated recently.
In order to discuss the origins of these three components further,
we would need further CO and/or OH-maser observations 
to distinguish their spatial distributions.

\subsubsection{Emission Morphology}\label{sec:image}

Here,
we discuss the spatial distribution of elements
with a combination of spatial and spectral analysis.
The ejecta component has already been identified with ejecta
heated by the reverse shock.
Specifically, the spectral analysis suggests that the Fe-K emission in Figure~\ref{fig:CSimage}
corresponds to ejecta
 (Figure~\ref{fig:n103b_spectra}).
Thus, the Fe-K image provides a good indication of
the distribution of the X-ray emitting ejecta.
The line emissions for elements from Si to Ca also correspond to
ejecta based on the spectral analysis,
although there is also Si emission from ISM.
In Figure~\ref{fig:CSimage},
the bright spots in the images for elements from Si to Ca are located to the southwest,
which is rather similar to the case of Fe-K.
This morphology supports our interpretation about the ejecta emission.
In contrast, in Figure~\ref{fig:CSimage} the Fe-K emission is located
slightly inward compared with the emissions of lighter elements.
The layered structure of ejecta could be preserved after a supernova explosion.
Similar ejecta distributions have also been reported
for the Tycho SNR observed with Suzaku/XIS based
on its Doppler motion \citep{2010ApJ...725..894H}.

In the ISM O and Fe-L emission images in Figure~\ref{fig:CSimage}, 
it appears as though they do not overlap with the ejecta emission.
This morphology also supports our interpretation about ISM emission.
The 0.5--0.9~keV band image support it again.
In addition, the ISM temperatures of $\sim$0.32, $\sim$0.55,
and $\sim$ 0.96~keV are significantly lower than the ejecta temperature of $\sim$3.96~keV.
These results indicate that the ejecta and ISM components are not
dynamically connected (see also \citet{2002A&A...392..955V}).
This result supports our discussion in \S~\ref{sec:abs_abundance}.

\begin{figure}
  \begin{center}
   \FigureFile(80mm,80mm)
   {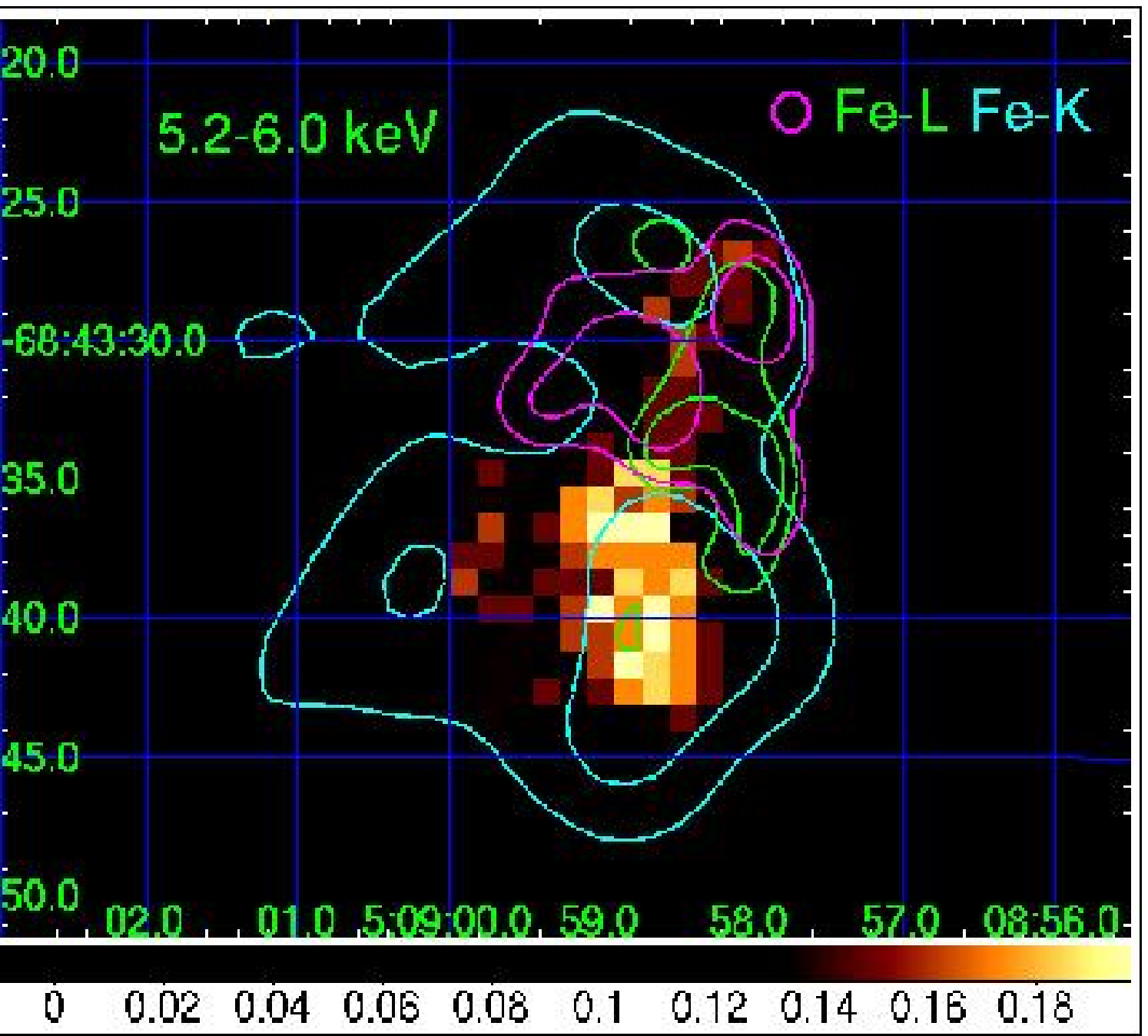}
  \end{center}
  \caption{Chandra/ACIS image of N103B in the 5.2--6.0~keV band.
The image is smoothed with a tophat function with $\sigma = 5$,
and a bin size of 2.5".
The overlapping contours represent
1.7 and 2.3~counts\,px$^{-1}$ for O, 3.5 (magenta),
4.0~counts\,px$^{-1}$ for Fe-L (green),
and 0.2 and 0.4~counts\,px$^{-1}$ for Fe-K (cyan).
The coordinates are in the equatorial coordinate system for epoch 2000.}
 \label{fig:high_energy_continuum}
\end{figure}

Figure~\ref{fig:high_energy_continuum} shows a Chandra/ACIS image
in the 5.2--6.0~keV band.
The overlapping contours represent O (magenta), Fe-L (green),
and Fe-K (cyan), respectively.
The high-energy continuum is located away from the ISM component
represented by O and Fe-L, 
and corresponds to the bright spot for Fe-K,
suggesting that
the origin of the high-energy continuum is the ejecta.

\subsection{Discussion on Progenitor of N103B}

In this section,
we discuss the progenitor of N103B.
In many cases, it is difficult to determine
whether the progenitor is of Type Ia or II.
Although the detection of a light echo spectrum and the presence of a pulsar
are powerful tools for determining the progenitor type,
in many cases we have to rely on abundance patterns,
heavy element masses, and/or the surrounding environment
to determine the progenitor type.
We discuss on the progenitor of N103B
with the spectral and imaging analysis results.


\subsubsection{Progenitor Type}\label{sec:method1}

The supernova explosion type is essentially classified
by the early-time spectra through optical observation.
Based on the optical spectra, Type I and Type II are defined
by the lack and presence of hydrogen structures, respectively \citep{1997ARA&A..35..309F}.
These spectral features reflect the envelope of the star
before the supernova explosion.
In essence, optical observations should indicate whether
the ejecta are composed of
little hydrogen and a lot of metal (Type I) or a lot of hydrogen and little metal (Type II).
These two types of ejecta are
referred to as pure metal plasma and H-dominated plasma, respectively.

Figure~\ref{fig:n103b_spectra} shows
the best-fit spectra assuming (a) H-dominated plasma
and (b) pure metal plasma.
We can see that the origin of the 5--6~keV band continuum
is different depending on the plasma type, where
the origin is the ejecta for the H-dominated plasma model
and the high-temperature ISM component in the pure metal plasma model.
However, we cannot determine which model is correct
on the basis of spectral analysis alone since the confidence intervals for the two models are comparable.
We therefore utilize Chandra/ACIS images to determine the correct model.
The spatial distribution of the high-energy continuum most likely corresponds to ejecta since it is similar
to a distribution for ejecta.
Combining spectral analysis based on Suzaku/XIS data
with image analysis based on Chandra/ACIS data,
we concluded that the progenitor of this remnant consisted of 
H-dominated plasma, in other words, it was a Type II progenitor.

A recent study \citep{2011ApJ...732..114L} favors a Type Ia progenitor for N103B.
However, our results reveal some uncertainty in that study.
For example,
\citet{2011ApJ...732..114L} reported that the progenitor might be Type Ia
on the basis of statistical image analysis of the morphologies from Si~XIII and 0.5--2.1~keV emissions.
However, applying  that method to the present sample might be problematic
since the Si~XIII and 0.5--2.1~keV emissions are
a combination of emissions from ISM and ejecta.
The fact that N103B is associated with
recent star formation \citep{2009ApJ...700..727B}
may make it difficult to determine the progenitor type of N103B, and thus
further studies are needed in order to reach an unambiguous conclusion.

With our new method,
we can determine the progenitor type by using only observational data,
without the need for a nucleosynthesis model.
This method can help to distinguish between pure metal plasma (Type I)
and H-dominated plasma (Type II) on the basis of detailed X-ray observations.
The advantage of this method is the lower uncertainty
compared with other methods (e.g., using plasma volume and a nucleosynthesis model).
However, the disadvantages of this method are
that it fails to distinguish between the Type Ia and Type Ib/c subclasses
Type I
and that it cannot provide an estimate of the progenitor mass.
To obtain this information, we have to rely on other well-known methods.

\subsubsection{Heavy Element Masses and Abundance Patterns}\label{sec:method2}

In the previous section,
we showed that the progenitor was most likely Type II
although there is still some uncertainty.
Here,
we estimate the progenitor mass
by comparing the results of a nucleosynthesis model with observational data
under the assumption that N103B has a Type II progenitor. 

To calculate the hydrogen number density,
we need to know the ejecta volume heated by the reverse shock.
We therefore assumed a sphere with a radius of $\sim4"\simeq0.93$~pc
for the ejecta volume ($V_{\rm{ejecta}}$),
which was estimated from the southwest Fe-K knot
shown in Figure~\ref{fig:CSimage}.
From this, $V_{\rm{ejecta}}$ was estimated to be
$\sim9.9\times10^{55}~\rm{cm^{-3}}$
on the assumption that the plasma was contained in the uniform sphere and not in the shell.
With the aid of the observed emission measure
(1.97$^{+0.03}_{-0.01}\times10^{58}~\rm{cm^{-3}}$),
$V_{\rm{ejecta}}$ and the volume filling factor ($f$),
we calculated $n_{\rm{H}}$ by taking into account $n_{\rm{e}} = 1.2n_{\rm{H}}$
for fully ionized solar abundance plasma \citep{1989GeCoA..53..197A}
since the ejecta were reproduced by the H-dominated plasma model.
In this case, $n_{\rm{H}}$ was estimated to be
$\sim12.87^{+0.10}_{-0.03}f^{-0.5}~\rm{cm^{-3}}$.
 
The number density of heavy element i ($n_{\rm{i}}$) is given by
\begin{equation}
\label{eq:densiy}
n_{\rm{i}} = A_{\rm{i\odot}}  A_{\rm{i}} n_{\rm{H}},
\end{equation}
where $A_{\rm{i\odot}}$ and $A_{\rm{i}}$ are
the solar abundance \citep{1989GeCoA..53..197A}
and the observed abundance relative to the solar abundance, respectively.
We assumed that the proton mass ($m_{\rm{p}}$) is equal to the neutron mass,
and the electron mass was ignored.
Then, the mass of heavy element i ($M_{\rm{i}}$) is given by
\begin{equation}
\label{eq:mass}
M_{\rm{i}} =  \eta_{\rm{i}} m_{\rm{p}} n_{\rm{i}} V_{\rm{i}},
\end{equation}
where $\eta_{\rm{i}}$ and $V_{\rm{i}}$ are the mass number and
ejecta volume of heavy element i, respectively. 

We assumed that $V_{\rm{i}}$ for each element was equal to
$V_{\rm{ejecta}}$,
since the images for elements from Si to Fe-K were similar to that of Fe-K
(Figure~\ref{fig:CSimage}).
We also assumed that $V_{\rm{i}}$ for Ne and Mg was equal to $V_{\rm{ejecta}}$,
although it is difficult to estimate $V_{\rm{i}}$ for Ne and Mg
due to the ISM emission.
Then, combining Equations~\ref{eq:densiy} and~\ref{eq:mass},
we can obtain the heavy element masses. 

The obtained number density and mass of each element are
listed in Table~\ref{table:ejecta_properties}.
For comparison, we also listed the heavy element masses
calculated with the nucleosynthesis model for 13~M$_\odot$
\citep{1997NuPhA.616...79N}
and Type Ia W7 \citep{1999ApJS..125..439I}, respectively.
The isotope masses calculated from the nucleosynthesis model
for the same element were summed. 
The observed masses were significantly lower compared to
those calculated with the nucleosynthesis model.
The ejecta morphology has a strongly asymmetric nature,
and it appears that the eastern part of N103B may start to be heated
only recently
by the reverse shock
because of non-uniform surrounding matter and/or asymmetric explosion.
If the ejecta is not uniformly heated,
the observed mass is lower than that calculated with the nucleosynthesis model. 
\citet{2002A&A...392..955V} reported the masses of Si and Fe
to be 0.032 and 0.033~M$_\odot$, respectively.
These heavy element masses are close to
the observed values, with a deviation of up to  a factor of 3. 

\begin{table*}
  \caption{Number densities and masses of heavy elements.}
  \label{table:ejecta_properties}
  \begin{center}
    \begin{tabular}{crrcc}
      \hline
      Element & \multicolumn{1}{c}{$n_{\rm{i}}$} & \multicolumn{1}{c}{$M_{\rm{i}}$} & 13~M$_\circ$ & W7\\
                      & \multicolumn{1}{c}{($10^{-3}$~cm$^{-3}$)} & \multicolumn{3}{c}{($10^{-3}$~M$_\odot$)}  \\\hline
      Ne & $<0.1f^{-0.5}$                      & $<0.2f^{+0.5}$                        & 22   & 4.5  \\
      Mg & 1.1 (1.0--1.2)$f^{-0.5}$       & 2.2 (2.0--2.4)$f^{+0.5}$        & 12  & 8.6   \\
      Si   & 4.4 (4.3--4.6)$f^{-0.5}$       & 10.3 (10.0--10.7)$f^{+0.5}$ & 70  & 157  \\
      S    & 3.6 (3.5--3.7)$f^{-0.5}$       & 9.6 (9.3--9.9)$f^{+0.5}$        & 17  &   87  \\
      Ca & 0.9 (0.8--1.0)$f^{-0.5}$        & 3.0 (2.7--3.3)$f^{+0.5}$        & 2.7  &  12  \\
      Fe  & 2.52 (2.46--2.59)$f^{-0.5}$ & 10.9 (10.7--11.2)$f^{+0.5}$ & 157 & 749 \\
      Ni   & 0.42 (0.39--0.47)$f^{-0.5}$ & 1.9 (1.8--2.2)$f^{+0.5}$        & 11   & 126 \\
      \hline
    \end{tabular}
  \end{center}
\end{table*}

We estimated the progenitor mass by comparing the abundance pattern
calculated with the nucleosynthesis model
with that obtained from spectral analysis.
With Equations~\ref{eq:densiy} and \ref{eq:mass},
the abundance pattern calculated with the nucleosynthesis model
relative to the solar abundance of elements i and j
($\equiv \frac{A_{\rm{i}}}{A_{\rm{j}}}$) is given by 
\begin{equation}
\label{eq:expect}
\frac{\rm{A_{\rm{i}}}}{\rm{A_{\rm{j}}}} = \frac{ \frac{n_{\rm{i}}}{n_{\rm{H}}A_{\rm{i\odot}}} }{ \frac{n_{\rm{j}}}{n_{\rm{H}}A_{\rm{i\odot}}} } = \frac{ n_{\rm{i}}A_{\rm{j\odot}} }{ n_{\rm{j}}A_{\rm{i\odot}} } = \frac{ \frac{M_{\rm{i}}}{\eta_{\rm{i}}m_{\rm{p}}V_{\rm{i}}} A_{\rm{j}\odot} }{ \frac{M_{\rm{j}}}{\eta_{\rm{j}}m_{\rm{p}}V_{\rm{j}}} A_{\rm{i}\odot} } = \frac{ \eta_{\rm{j}} M_{\rm{i}} V_{\rm{j}} A_{\rm{j\odot}} }{ \eta_{\rm{i}} M_{\rm{j}} V_{\rm{i}} A_{\rm{i\odot}}}.
\end{equation}
To calculate the abundance pattern,
the isotope masses of each element were summed,
and the averaged isotope mass was used as the mass number.

\begin{figure}
 \begin{center}
   \FigureFile(80mm,80mm)
   {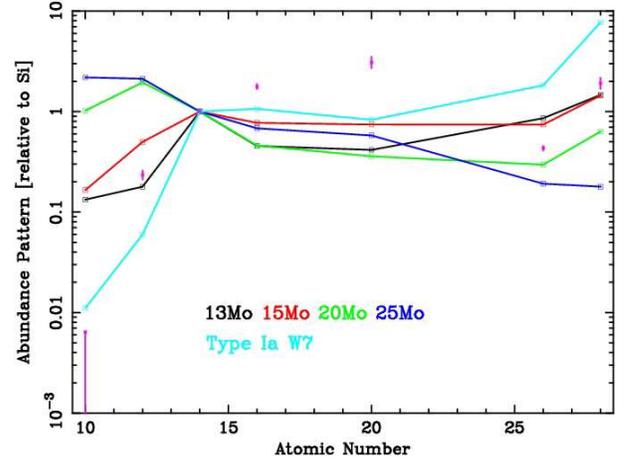}
  \end{center}
  \caption{Atomic number plotted against the calculated and observed abundance patterns
relative to Si and the solar abundance.
Black, red, green, and blue points represent
the abundance patterns for 13, 15, 20, and 25~M$_\circ$
assuming a CC progenitor \citep{1997NuPhA.616...79N}.
The cyan points represent the results for Type Ia W7
\citep{1999ApJS..125..439I}.
The purple data points represent the observed abundances pattern.}
 \label{fig:abs_pattern}
\end{figure}

Assuming that heavy elements i and j have the same volume,
the abundance pattern can be calculated from Equation~\ref{eq:expect}.
The data points plotted in cyan in Figure~\ref{fig:abs_pattern}
represent the abundance pattern relative to that of Si and the solar abundance.
For comparison, we also plotted
the ratio predicted from the nucleosynthesis models for 13~M$_\odot$ (black),
15~M$_\odot$ (red), 20~M$_\odot$ (green), and 25~M$_\odot$ (blue)
\citep{1997NuPhA.616...79N}
and Type Ia W7 (cyan) \citep{1999ApJS..125..439I}.
Our results show that 

As shown in Figure~\ref{fig:abs_pattern},
the relative abundance of Mg is lower than Si,
indicating the progenitor is a rather low-mass star.
Fe and Ni abundance is rather high,
which also indicates the low-mass progenitor.
With these results,
we estimated the progenitor mass as 13~M$_\odot$.
Note that the volume of Mg and Ni ejecta can be different
from those of other elements, such as Si and Fe.
$\frac{\rm{Mg}}{\rm{Si}}$ and $\frac{\rm{Ni}}{\rm{Si}}$ are therefore
associated with a large uncertainty.

\subsection{Propagation of Reverse Shock in Iron-rich Ejecta}

In \S\ref{sec:spec:iron},
we showed that
the plasma emitting the Fe-K line has a relatively low degree of ionization
compared with plasma showing emission from other elements.
A similar result is also reported in 
\citet{2002A&A...392..955V}.
Fe-K line emission originates from ejecta heated by the reverse shock,
and thus the reverse shock might be propagating in Fe-rich ejecta at present.

\subsection{Density, Total Energy, Swept-up Mass, and Age}

\begin{table*}
  \caption{Input and derived properties in the case of H-dominated plasma.}
  \label{table:ism_properties}
  \begin{center}
    \begin{tabular}{lrrr}
      \hline
 	Property                & \multicolumn{3}{c}{ISM components} \\
	                               & \multicolumn{1}{c}{Low-$T_{\rm{e}}$} & \multicolumn{1}{c}{Middle-$T_{\rm{e}}$} & \multicolumn{1}{c}{High-$T_{\rm{e}}$} \\
      \hline
     	$kT_{\rm{e}}$ (keV) \,.........                     & 0.319 (0.314--0.326) & 0.547 (0.541--0.552) & 0.962 (0.957--0.967) \\
	$R$ (pc) \,..............                                    & \multicolumn{1}{c}{3.49} & \multicolumn{1}{c}{3.49} & \multicolumn{1}{c}{3.49} \\
	$V$ ($10^{57}~\rm{cm^{-3}}$) ..            & \multicolumn{1}{c}{1.42} & \multicolumn{1}{c}{1.42} & \multicolumn{1}{c}{1.42} \\
	$n_{\rm{e}}$ ($\rm{cm^{-3}}$) .........      &10.0 (9.7--13.5)$f^{-0.5}$ & 14.0 (13.9--14.2)$f^{-0.5}$ & 14.8 (14.7--15.0)$f^{-0.5}$ \\
	$n_0$ ($\rm{cm^{-3}}$) .........                 &     2.1 (2.0--2.8)$f^{-0.5}$ & 2.92 (2.89--2.95)$f^{-0.5}$ & 3.09 (3.05--3.13)$f^{-0.5}$ \\
	$E_{\rm{t}}$ ($10^{49}~\rm{erg}$) \,.... & 2.2 (2.1--3.0)$f^{+0.5}$ & 5.2 (5.1--5.3)$f^{+0.5}$ & 9.7 (9.6--9.9)$f^{+0.5}$ \\
	$M_{\rm{swept}}$ (M$_\odot$) \,....       & 9.9 (9.6--13.4)$f^{+0.5}$ & 14.0 (13.8--14.1)$f^{+0.5}$ & 14.8 (14.9--14.6)$f^{+0.5}$ \\
	$t_{\rm{ion}}$ (yr) \,............                      & 100 (170--60)$f^{+0.5}$ & $>$1700$f^{+0.5}$ & 166 (172--160)$f^{+0.5}$ \\ 
      \hline
    \end{tabular}
  \end{center}
\end{table*}

Owing to the successful separation of the three ISM components
from the single ejecta component,
we can estimate the environment around N103B.
In this section, we estimate the pre-shock ($n_{\rm{0}}$)
and post-shock ($n_{\rm{e}}$) number density,
the thermal energy ($E_{\rm{t}}$), the swept-up mass ($M_{\rm{swept}}$),
and the ionization age of plasma ($t_{\rm{ion}}$)
to gain insight into the surrounding environment and the plasma nature for this remnant. 

With the aid of the emission measure obtained from the spectral analysis,
the volume of the plasma ($V$), and the volume filling factor ($f$),
we calculated $n_{\rm{e}}$ by considering $n_{\rm{e}} = 1.2n_{\rm{H}}$
for fully ionized solar abundance plasma \citep{1989GeCoA..53..197A}.
Under the assumption of strong shock and ideal gas,
the pre-shock number density is obtained
from $n_{\rm{0}} = \frac{n_{\rm{H}}}{4}$
(known as the {\it Rankine-Hugoniot} relation),
although recent studies on efficient cosmic ray acceleration
in the SNR shocks have demonstrated that a higher compression ratio is required
(e.g., \citet{bamba2005}).
The swept-up mass is given by $M_{\rm{swept}} = n_{\rm{H}}m_{\rm{p}}Vf$, and
the total thermal energy is given by $E_{\rm{t}} = 3n_{\rm{e}}kT_{\rm{e}}Vf$,
where $k$ and $T_{\rm{e}}$ are the Boltzmann constant
and the absolute temperature of the plasma, respectively.
Electrons and ions are assumed to be in thermal equilibrium.
The ionization age of plasma is calculated from $n_{\rm{e}}$ and
$n_{\rm{e}}t_{\rm{ion}}$ obtained in the spectral analysis. 

In order to estimate the plasma number density,
we need to know the volume of the plasma, which can be derived from X-ray images.
In \S~\ref{sec:image}, we have already shown that the emission from ISM occupies
the western region of the remnant.
We therefore assumed that the plasma was compressed
in the shell of the hemisphere with a radius and a thickness of
$\sim$15'' and $\sim$1.5'', respectively.
The shell thickness was taken from \citet{2002A&A...392..955V}
and \citet{1992sswi.book.....L}.
Assuming a distance to LMC of 48~kpc \citep{2006ApJ...652.1133M},
the radius and the thickness of the shell were estimated
to be $\sim$3.49 and $\sim$0.35~pc, respectively.
As a result,
the volume of the plasma was estimated to be $V \simeq 1.42\times10^{57}~\rm{cm^{3}}$.
The obtained parameters are summarized in Table~\ref{table:ism_properties}.
In the ISM images for elements such as O and Fe-K (Figure~\ref{fig:CSimage}),
the morphology of this remnant is highly anisotropic.
We emulated this effect by changing the volume filling factor $f$
between 0.1 and 1.0
in order to estimate the number density, thermal energy,
swept-up mass, and ionization age of the plasma. 

Using the abovementioned values for the volume filling factor,
the plasma number density of the low-, middle-, and high-temperature components
were estimated to be $\sim$10--32, $\sim$14--44, and $\sim$15--47~cm$^{-3}$,
respectively.
These plasma number densities were roughly consistent with
the results obtained from XMM-Newton observations ($\sim$7--25~cm$^{-3}$: \citep{2002A&A...392..955V}).
Then, the pre-shock (ambient) number densities
of the low-, middle-, and high-temperature components were estimated to be
$\sim$2--7, $\sim$3--7, and $\sim$3--10~cm$^{-3}$.
These relatively high ambient number densities imply
a physical connection to the HII region DEM L84 \citep{2001ApJS..136..119D}
and the young rich cluster NGC 1850 \citep{1988AJ.....96.1874C}
near this remnant.
The complex images obtained with Chandra/ACIS might reflect such a complex environment.
We also confirmed that the supernova explosion for this remnant
occurred in a high-density region with an ambient density
of $\sim$2--10~cm$^{-3}$.

The ionization ages of the low-, middle-, and high-temperature plasma
components
were estimated to be $\sim$30--100, $>$1700, and $\sim$50--170~yr,
respectively,
using the same assumption for the filling factor.
These values are close to those
obtained from XMM-Newton observations \citep{2002A&A...392..955V}.
\citet{2005Natur.438.1132R} estimated the age of this remnant to be 860~yr
based on the light echo.
The age for the middle-temperature ISM component obtained in this study is consistent with
this estimation.
In contrast, the low- and high-temperature ISM components
were estimated to be rather young at $\sim$30--170~yr, suggesting that
these two components might have been heated recently by the forward shock. 

The total thermal energy was defined
by $E_{\rm{t}}^{\rm{total}} = E_{\rm{t}}^{\rm{low}} + E_{\rm{t}}^{\rm{middle}} + E_{\rm{t}}^{\rm{high}}$,
where $E_{\rm{t}}^{\rm{low}}$, $E_{\rm{t}}^{\rm{middle}}$ and
$E_{\rm{tl}}^{\rm{high}}$ denote the thermal energies of the low-, middle-,
and high-temperature components, respectively.
$E_{\rm{t}}^{\rm{total}}$ was estimated to be
$\sim1.7\times10^{50}f^{+0.5}~\rm{erg}$.
Under the same assumption for the filling factor,
the total thermal energy was estimated
to be $\sim0.5$--$1.7\times10^{50}~\rm{erg}$,
which is close to that of a general SNR
($\sim$0.5--7$\times10^{51}$~erg: \citep{1998ApJ...505..732H}).
$M_{\rm{swept}}^{\rm{total}}$ was obtained
by $M_{\rm{swept}}^{\rm{low}} + M_{\rm{swept}}^{\rm{middle}} + M_{\rm{swept}}^{\rm{high}}$,
where $M_{\rm{swept}}^{\rm{low}}$, $M_{\rm{swept}}^{\rm{middle}}$ and $M_{\rm{swept}}^{\rm{high}}$
denote the swept-up masses of the low-,
middle- and high-temperature components, respectively.
In this case, $M_{\rm{swept}}^{\rm{total}}$ was estimated to be 38.7~M$_\odot$.
Assuming the same volume filling factor,
the swept-up mass was estimated to be 12.2--38.7~M$_\odot$.
This high value indicates that
the three-temperature plasma components essentially consist of
swept-up ambient material.

\section{Summary}\label{sec:sum}

We observed the supernova remnant N103B in the LMC
with Suzaku and Chandra.
In the spectral analysis conducted using Suzaku/XIS data,
the diffuse thermal emission was closely reproduced
with three ISM components with temperatures of $\sim$0.32, $\sim$0.55,
and $\sim$0.96~keV and the one ejecta component with a temperature of $\sim$3.96~keV.
The abundances of the ISM components were close to the LMC averages
\citep{1992ApJ...384..508R},
and therefore we concluded that these components were heated by the forward shock.
The ejecta component was overabundant in heavy elements such as
Mg, Si, S, Ca, Fe, and Ni.
Our interpretation is that this component originates from ejecta.
The unprecedentedly high quality of the data
allowed us to distinguish between ISM and ejecta emissions for the first time in a spectral analysis. 

We also analyzed images obtained with Chandra/ACIS
to examine the validity of the spectral analysis conducted with Suzaku/XIS data.
We revealed that the distributions of elements from Si to Fe-K were similar
to each other, although Fe-K was located slightly inward
in comparison with lighter elements.
This morphology supports our interpretation about ejecta-based emission.
ISM emission (represented by O and Fe-L) is distributed
differently than the ejecta emission.
This morphology also supports our interpretation
about ISM emissions.
In addition, the ISM temperatures of $\sim$0.32, $\sim$0.55,
and $\sim$0.96~keV were significantly lower
compared to the ejecta temperature of $\sim$3.92~keV.
These results indicate
that the ejecta and ISM emissions were not dynamically connected
\citep{2002A&A...392..955V}. 

Combining spectral analysis based on Suzaku/XIS data
and the high-energy continuum
in the 5.2--6.0~keV band obtained with Chandra/ACIS,
we showed the indication that the progenitor was most likely Type II.
Assuming this progenitor type
the progenitor mass was estimated to be 13~M$_\odot$
based on the abundance patterns of Mg, Fe, and Ni relative to Si. 

Based on the identification of the three ISM components
from the single ejecta component,
we also gained insight into the surrounding environment of N103B.
The ambient number density for the three ISM components
was estimated to be $\sim$2--10~cm$^{-3}$.
This relatively high ambient number density implies
a physical connection with the HII region DEM L84 \citep{2001ApJS..136..119D}
and the young rich cluster NGC 1850 \citep{1988AJ.....96.1874C}
near this remnant.
Thus, we revealed that the supernova explosion which gave rise to N103B occurred
in a high-density region.

\section*{Acknowledgements}

We thank the anonymous referee for fruitful comments.
We also thank Y. Maeda, H. Yamaguchi, H. Yoshitake, I. Mitsuishi,
and T. Hayashi for many useful discussions and comments.
We thank all members of the Suzaku team
for their contributions to instrument preparation,
spacecraft operation, software development,
and in-orbit instrument calibration.
We made use of archive data obtained from the {\it Chandra X-ray Center}, and
we are grateful to the Chandra/ACIS team.
This work was supported in part by a Grant-in-Aid for Scientific Research No.~22684012 (A.~B) provided
by the Japanese Ministry of Education, Culture, Sports, Science and Technology.



\begin{thebibliography}{}

\bibitem[Anders \& Grevesse(1989)]{1989GeCoA..53..197A}
Anders, E., \& Grevesse, N.\ 1989, \gca, 53, 197 

\bibitem[Badenes et al.(2009)]{2009ApJ...700..727B} Badenes, C., Harris, 
J., Zaritsky, D., \& Prieto, J.~L.\ 2009, \apj, 700, 727 

\bibitem[Bamba et al.(2005)]{bamba2005} Bamba, A., Yamazaki, R., 
Yoshida, T., Terasawa, T., \& Koyama, K.\ 2005, \apj, 621, 793 

\bibitem[Bamba et al.(2008)]{2008PASJ...60S.153B} Bamba, A., et al.\ 2008, 
\pasj, 60, 153 

\bibitem[Borkowski et al.(2001)]{2001ApJ...548..820B} Borkowski, K.~J., 
Lyerly, W.~J., \& Reynolds, S.~P.\ 2001, \apj, 548, 820 

\bibitem[Borkowski et al.(1994)]{1994ApJ...429..710B} Borkowski, K.~J., 
Sarazin, C.~L., \& Blondin, J.~M.\ 1994, \apj, 429, 710 

\bibitem[Chu \& Kennicutt(1988)]{1988AJ.....96.1874C}
Chu, Y.-H., \& Kennicutt, R.~C., Jr.\ 1988, \aj, 96, 1874 

\bibitem[Chu et al.(2000)]{2000AJ....119.2242C} Chu, Y.-H., Kim, S., 
Points, S.~D., Petre, R., \& Snowden, S.~L.\ 2000, \aj, 119, 2242 

\bibitem[Dickey \& Lockman(1990)]{1990ARA&A..28..215D}
Dickey, J.~M., \& Lockman, F.~J.\ 1990, \araa, 28, 215 

\bibitem[Dunne et al.(2001)]{2001ApJS..136..119D} Dunne, B.~C., Points, 
S.~D., \& Chu, Y.-H.\ 2001, \apjs, 136, 119 

\bibitem[Filippenko(1997)]{1997ARA&A..35..309F}
Filippenko, A.~V.\ 1997, \araa, 35, 309 

\bibitem[Glatt et 
al.(2010)]{2010A&A...517A..50G} Glatt, K., Grebel, E.~K., \& Koch, A.\ 2010, \aap, 517, A50 

\bibitem[Hamilton et al.(1983)]{1983ApJS...51..115H} Hamilton, A.~J.~S., 
Chevalier, R.~A., \& Sarazin, C.~L.\ 1983, \apjs, 51, 115 

\bibitem[Hayato et al.(2010)]{2010ApJ...725..894H} Hayato, A., et al.\ 
2010, \apj, 725, 894 

\bibitem[Hughes et al.(1995)]{1995ApJ...444L..81H} Hughes, J.~P., et al.\ 
1995, \apjl, 444, L81 

\bibitem[Hughes et al.(1998)]{1998ApJ...505..732H} Hughes, J.~P., Hayashi, 
I., \& Koyama, K.\ 1998, \apj, 505, 732 

\bibitem[Hughes et al.(2000)]{2000ApJ...543L..61H} Hughes, J.~P., Rakowski, 
C.~E., \& Decourchelle, A.\ 2000, \apjl, 543, L61 

\bibitem[Hwang et al.(2000)]{2000ApJ...537L.119H} Hwang, U., Holt, S.~S., 
\& Petre, R.\ 2000, \apjl, 537, L119 

\bibitem[Ishisaki et al.(2007)]{2007PASJ...59S.113I} Ishisaki, Y., et al.\ 
2007, \pasj, 59, 113 

\bibitem[Iwamoto et al.(1999)]{1999ApJS..125..439I} Iwamoto, K., Brachwitz, 
F., Nomoto, K., Kishimoto, N., Umeda, H., Hix, W.~R., 
\& Thielemann, F.-K.\ 1999, \apjs, 125, 439 


\bibitem[Keller 
\& Wood(2006)]{2006ApJ...642..834K} Keller, S.~C., \& Wood, P.~R.\ 2006, \apj, 642, 834 

\bibitem[Kosenko et al.(2010)]{2010A&A...519A..11K}
Kosenko, D., Helder, E.~A., \& Vink, J.\ 2010, \aap, 519, A11 

\bibitem[Koyama et al.(2007)]{2007PASJ...59S..23K} Koyama, K., et al.\ 
2007, \pasj, 59, 23 

\bibitem[Lewis et al.(2003)]{2003ApJ...582..770L} Lewis, K.~T., Burrows, 
D.~N., Hughes, J.~P., Slane, P.~O., Garmire, G.~P., 
\& Nousek, J.~A.\ 2003, \apj, 582, 770 

\bibitem[Liedahl et al.(1995)]{1995ApJ...438L.115L} Liedahl, D.~A., 
Osterheld, A.~L., \& Goldstein, W.~H.\ 1995, \apjl, 438, L115 

\bibitem[Lopez et al.(2011)]{2011ApJ...732..114L} Lopez, L.~A., 
Ramirez-Ruiz, E., Huppenkothen, D., Badenes, C., 
\& Pooley, D.~A.\ 2011, \apj, 732, 114 

\bibitem[Lozinskaya(1992)]{1992sswi.book.....L} Lozinskaya, T.~A.\ 1992, 
New York: American Institute of Physics, 1992,  

\bibitem[Macri et al.(2006)]{2006ApJ...652.1133M} Macri, L.~M., Stanek, 
K.~Z., Bersier, D., Greenhill, L.~J., \& Reid, M.~J.\ 2006, \apj, 652, 1133 

\bibitem[Mathewson et al.(1983)]{1983ApJS...51..345M} Mathewson, D.~S., 
Ford, V.~L., Dopita, M.~A., Tuohy, I.~R., Long, K.~S., 
\& Helfand, D.~J.\ 1983, \apjs, 51, 345 

\bibitem[Mitsuda et al.(2007)]{2007PASJ...59S...1M} Mitsuda, K., et al.\ 
2007, \pasj, 59, 1 

\bibitem[Nakajima et al.(2008)]{nakajima2008}
Nakajima, H.\ et al.\ 2008, \pasj, 60, S1


\bibitem[Nomoto et al.(1997)]{1997NuPhA.616...79N} Nomoto, K., Hashimoto, 
M., Tsujimoto, T., Thielemann, F.-K., Kishimoto, N., Kubo, Y., 
\& Nakasato, N.\ 1997, Nuclear Physics A, 616, 79 

\bibitem[Rest et al.(2005)]{2005Natur.438.1132R} Rest, A., et al.\ 2005, 
\nat, 438, 1132 

\bibitem[Russell 
\& Dopita(1992)]{1992ApJ...384..508R} Russell, S.~C., \& Dopita, M.~A.\ 1992, \apj, 384, 508 

\bibitem[Serlemitsos et al.(2007)]{2007PASJ...59S...9S} Serlemitsos, P.~J., 
et al.\ 2007, \pasj, 59, 9 

\bibitem[Someya et al.(2010)]{2010PASJ...62.1301S} Someya, K., Bamba, A., 
\& Ishida, M.\ 2010, \pasj, 62, 1301 

\bibitem[Uchiyama et al.(2009)]{uchiyama2009}
Uchiyama, H., et al.\ 2009, \pasj, 61, 9 


\bibitem[van der Heyden et al.(2002)]{2002A&A...392..955V} van der Heyden, K.~J., Behar, E., Vink, J., Rasmussen, A.~P., Kaastra, J.~S., Bleeker, J.~A.~M., Kahn, S.~M., \& Mewe, R.\ 2002, \aap, 392, 955 

\bibitem[Vink et 
al.(1996)]{1996A&A...307L..41V} Vink, J., Kaastra, J.~S., \& Bleeker, J.~A.~M.\ 1996, \aap, 307, L41 

\bibitem[Yamaguchi et al.(2008)]{2008PASJ...60S.141Y} Yamaguchi, H., et 
al.\ 2008, \pasj, 60, 141 

\bibitem[Yang et al.(2013)]{2013ApJ...766...44Y} Yang, X.~J., Tsunemi, H., 
Lu, F.~J., et al.\ 2013, \apj, 766, 44 

\end{thebibliography}
\end{document}